\pdfoutput=1 

\documentclass[a4paper, 11pt, final]{article}

\usepackage{amssymb} 
\usepackage{amsthm} 
\usepackage{amsmath,empheq}
\usepackage{bbm}

\DeclareMathAlphabet{\mathcal}{OMS}{cmsy}{m}{n}
\DeclareMathAlphabet\mathbfcal{OMS}{cmsy}{b}{n}

\DeclareFontFamily{U}{dutchcal}{\skewchar\font=45 }
\DeclareFontShape{U}{dutchcal}{m}{n}{<-> s*[1.0] dutchcal-r}{}
\DeclareFontShape{U}{dutchcal}{b}{n}{<-> s*[1.0] dutchcal-b}{}
\DeclareMathAlphabet{\mathcald}{U}{dutchcal}{m}{n}
\SetMathAlphabet{\mathcald}{bold}{U}{dutchcal}{b}{n}
\DeclareMathAlphabet\mathcalz{T1}{pzc}{mb}{it}

\usepackage[nice]{nicefrac} 
\usepackage{siunitx} 

\usepackage{newtxtext, newtxmath} 

\usepackage{anyfontsize}
\usepackage[utf8]{inputenc}
\usepackage[T1]{fontenc}
\usepackage{microtype} 

\providecommand{\JEL}[1]{\textit{\textbf{JEL: }} #1}
\providecommand{\keywords}[1]{\textbf{\textit{Keywords--- }} #1}

\usepackage{setspace} 

\usepackage{parskip} 
\usepackage{etoolbox}
\usepackage{titlesec}
\titleformat{\section}{\normalfont\Large\bfseries}{\thesection}{1em}{}
\titleformat{\subsection}{\normalfont\large\bfseries}{\thesubsection.}{1em}{}
\titleformat{\subsubsection}{\normalfont\normalsize\itshape}{\thesubsubsection.}{1em}{}

\usepackage{abstract}

\renewenvironment{abstract}
 {\normalfont
  \begin{center}
  \bfseries \abstractname\vspace{-.5em}\vspace{0pt}
  \end{center}
  \list{}{
    \setlength{\leftmargin}{0cm}%
    \setlength{\rightmargin}{\leftmargin}%
  }%
  \item\relax}
 {\endlist}

\usepackage{authblk} 
\usepackage[bottom]{footmisc} 

\usepackage{multicol} 

\usepackage{enumitem} 

\usepackage{graphicx} 
\usepackage{float} 
\usepackage{subcaption}
\usepackage{afterpage} 

\usepackage{printlen}

\usepackage[labelfont=bf]{caption}
\usepackage{color, colortbl}
\definecolor{LightGray}{rgb}{0.93,0.914,0.914}    
\usepackage{soul} 
\usepackage{longtable,rotating} 
\usepackage{booktabs} 
\usepackage{multirow} 
\usepackage{arydshln} 
\usepackage{bigdelim} 

\usepackage[most]{tcolorbox}            

\PassOptionsToPackage{hyphens}{url} 

\newcommand*\rot{\rotatebox{90}}

\usepackage{fancyhdr}
\fancyheadoffset{0cm}
\makeatletter
\newcommand{\quickwordcount}[1]{
  \immediate\write18{texcount -quiet -incbib -sub=none -utf8 -1 -sum -merge -encoding=utf8 #1.tex > #1-words}%
  \immediate\openin\somefile=#1-words
  \read\somefile to \@@localdummy
  \immediate\closein\somefile
  \setcounter{wordcounter}{\@@localdummy}
  \@@localdummy
}
\makeatother

\usepackage{silence}
\usepackage[style=apa,backend=biber,natbib,hyperref]{biblatex}
\DeclareLanguageMapping{british}{british-apa}
\defbibenvironment{bibliography}{\enumerate}{\endenumerate}{\item}

\usepackage[colorlinks=true,allcolors=gray]{hyperref} 
\let\orgautoref\autoref

\renewcommand{\autoref}[1]
{%
\def\equationautorefname{Eq.}%
\def\sectionautorefname{Sec.}%
\def\subsectionautorefname{Subsec.}%
\def\figureautorefname{Fig.}%
\def\subfigureautorefname{Fig.}%
\orgautoref{#1}%
}

\usepackage[nameinlink,capitalise]{cleveref}

\usepackage{algorithm,algpseudocode}

\makeatletter
\newlength{\trianglerightwidth}
\algnewcommand{\LineCommentCont}[1]{\Statex \hskip\ALG@thistlm%
  \parbox[t]{\dimexpr\linewidth-\ALG@thistlm}
{\leftskip=\algorithmicindent
  \hangindent=\algorithmicindent 
  \hangafter=1%
  \strut\makebox[\algorithmicindent][c]{$\triangleright$}#1\strut}
  } 
\makeatother

\usepackage[left=1.8cm, right=1.8cm, bottom=2.5cm, top=2.5cm]{geometry}

\begin{document}


\renewcommand{\figureautorefname}{Fig.}
\onehalfspacing



\newcommand{\MainTitleText}{The TruEnd-procedure: Treating trailing zero-valued balances in credit data}

\title{\fontsize{20pt}{0pt}\selectfont\textbf{\MainTitleText
}}


\author[,a,b]{\large Arno Botha \thanks{ ORC iD: 0000-0002-1708-0153; email: \url{arno.spasie.botha@gmail.com}}}
\author[,a,b]{\large Tanja Verster \thanks{ ORC iD: 0000-0002-4711-6145; Corresponding author: \url{tanja.verster@nwu.ac.za}}}
\author[,a]{\large Roelinde Bester \thanks{ ORC iD: 0000-0002-1042-2579}}
\affil[a]{\footnotesize \textit{Centre for Business Mathematics and Informatics \& Unit for Data Science and Computing, North-West University, Potchefstroom, South Africa}}
\affil[b]{\footnotesize \textit{National Institute for Theoretical and Computational Sciences (NITheCS), Potchefstroom, South Africa}}
\renewcommand\Authands{, and }

    

\makeatletter
\renewcommand{\@maketitle}{
    \newpage
     \null
     \vskip 1em%
     \begin{center}%
      {\LARGE \@title \par
      	\@author \par
       }
     \end{center}%
     \par
 } 
 \makeatother
 
 \maketitle

{
    \setlength{\parindent}{0cm}
    \rule{1\columnwidth}{0.4pt}
    \begin{abstract}

    A novel procedure is presented for finding the true but latent endpoints within the repayment histories of individual loans. The monthly observations beyond these true endpoints are false, largely due to operational failures that delay account closure, thereby corrupting some loans. Detecting these false observations is difficult at scale since each affected loan history might have a different sequence of trailing zero (or very small) month-end balances. Identifying these trailing balances requires an exact definition of a "small balance", which our method informs.  We demonstrate this procedure and isolate the ideal small-balance definition using two different South African datasets. Evidently, corrupted loans are remarkably prevalent and have excess histories that are surprisingly long, which ruin the timing of risk events and compromise any subsequent time-to-event model, e.g., survival analysis. Having discarded these excess histories, we demonstrably improve the accuracy of both the predicted timing and severity of risk events, without materially impacting the portfolio. The resulting estimates of credit losses are lower and less biased, which augurs well for raising accurate credit impairments under IFRS 9. Our work therefore addresses a pernicious data error, which highlights the pivotal role of data preparation in producing credible forecasts of credit risk.

\end{abstract}
     
     
     \JEL{C41, C44, C61.}
    
    \rule{1\columnwidth}{0.4pt}
}

\noindent Word count (excluding front matter): 8089 

\subsection*{Disclosure of interest and declaration of funding}
\noindent This study is partially supported by the National Research Foundation of South Africa (Grant Number 126885). This work has no known conflicts of interest that may have influenced the outcome of this work. The authors also thank all anonymous referees and editors for their valuable contributions that have improved this work.



\newpage

\section{Introduction}
\label{sec:introduction}

Hailing from the biostatistical literature, survival analysis is a powerful class of techniques for modelling any time-to-event data across many disciplines; see \citet{singer1993time}, \citet{kleinbaum2012survival}, \citet{kartsonaki2016survival}, and \citet{schober2018survival}. Survival analysis examines the length of time until reaching some well-defined endpoint, should the event occur. By implication, survival models do not only predict the probability of an event occurring, but also its timing.
In quantitative finance, \citet{narain1992credit} originally demonstrated that a loan's \textit{probability of default} (or PD) is a function of time. \citet{banasik1999not} extended this work by developing Cox proportional hazards (PH) regression models using UK loans data, which compared favourably to cross-sectional logistic regression (LR) models. 
Another notable extension is \citet{stepanova2002survival}, who investigated Harrel's correlation method for finding time-dependent covariates, whereafter they compared various types of model diagnostics within the credit domain, e.g., Cox-Snell and Schoenfeld residuals. The authors also demonstrated a (more dynamic) time-sensitive binning procedure for Cox PH-models, which improves upon the otherwise static \textit{weight of evidence} (WOE) transform; itself commonly used within LR-models, as explained in \citet[\S 1.5]{thomas2009consumer}. 
In predicting the PD, other authors demonstrated that including time-dependent variables (especially macroeconomic variables such as interest/policy rates) into a Cox PH-model can improve the model fit and/or its predictive power, typically beyond that of a comparable LR-model; see \citet{bellotti2009macro}, \citet{crook2010dynamic}, \citet{bellotti2013}, and \citet{bellotti2014stresstesting}. 
Lastly, \citet{dirick2017time} compared various types of default survival models, including Cox PH with/without spline functions, mixture cure models, and accelerated failure time models.
Though certainly not exhaustive, this growing body of literature augurs well for the increasing use of survival analysis in predicting both the prevalence and timing of default.

The advent of the IFRS 9 accounting standard from \citet{ifrs9_2014} has generated greater impetus for more dynamic modelling such as survival analysis. In particular, IFRS 9 requires that a financial asset's value be comprehensively and regularly adjusted, based on the asset's \textit{expected credit loss} (ECL) over its lifetime; see \citet{novotny2016} and \citet{skoglund2017}. This ECL-estimation depends not only on the PD as risk parameter, but also on the \textit{Loss Given Default} (LGD), i.e., the fraction of the loan balance that was eventually lost, as measured at the point of default. 
However, this LGD-parameter is notoriously difficult to model due to the bimodality and asymmetry of the realised LGD-density, as discussed in \citet{schuermann2004we}, \citet{calabrese2010bank}, and \citet[pp.~271-314]{baesens2016credit}. These statistical complications arise naturally since the reality of resolving a defaulted loan can be complex, highly uncertain, lengthy, sensitive to the macroeconomic cycle, and even vary by jurisdiction; see \citet{schuermann2004we}, \citet[pp.~11-13]{finlay2010book}, \citet{gurtler2013LGD}, and \citet[pp.~23-26]{botha2021Proc}. 
Ultimately, it takes time to resolve a \textit{default spell} (or episode) that may end in write-off, whereupon all cash flows are discounted back to the default point in calculating the associated non-zero credit loss. Some default spells may resolve into a \textit{cured} outcome (with a zero-valued loss), while more recent spells are likely to be right-censored since collectors are still pursuing the resolution of these spells.
This framing suggests that survival analysis is a natural choice for modelling the LGD-parameter, particularly in predicting the time to either write-off or cure. In fact, a few studies have already explored some aspects hereof, though this topic is certainly still evolving; see \citet{witzany2012survival}, \citet{zhang2012comparisons}, \citet{wood2017addressing}, \citet{joubert2018making}, and \citet{larney2023modelling}.
Regarding write-off, IFRS 9 specifically calls for defining the timing of write-off as the point beyond which further loan recovery becomes questionable; see \citet[\S 5.4.4]{ifrs9_2014}. The consequences of a mis-timed write-off can therefore adversely affect the LGD-parameter, especially when using survival analysis as a modelling technique.

Despite methodological soundness, the inappropriate parametrisation of a survival model will inadvertently invite \textit{model risk}. In particular, \citet{dejongh_2017} define model risk as the adverse exposure resulting from a model that: 1) is conceptually flawed; 2) provides inaccurate output or predictions; 3) is used inappropriately; or 4) has a flawed implementation.
As such, a survival model that is trained on inaccurate or flawed data will render predictions that can disagree with reality quite substantially, thereby ruining subsequent decision-making as well as compromising the ECL-estimates for a bank's loss provision. Proper data preparation is itself paramount to producing any reliable statistical model, which can include many exercises such as data cleaning, feature engineering, data fusion, outlier detection, and missing value treatments, as discussed in \citet{james2013introduction}.
Moreover, the UK regulator has recently published five broad principles in managing model risk across banks, given the growing prevalence of using model outputs within decision-making; see \citet{pra2023PSmodelrisk} and \citet{pra2023SSmodelriskprinciples}. When using data during modelling, Principles 3.2 and 4.3 from \citet{pra2023SSmodelriskprinciples} call for verifying whether this data represents reality, as well as for assuring 
the quality and reliability of the same data. The South African regulator decrees similarly in Guidance Note 9 from \citet{sarb2022g9} that data should be both complete and truly relevant when using it for developing credit risk models. 
By pronouncing so pervasively on model risk, these regulators have only further affirmed the obvious importance of training models from data that is largely error-free.

\begin{figure}[ht!]
\centering\includegraphics[width=1\linewidth,height=0.12\textheight]{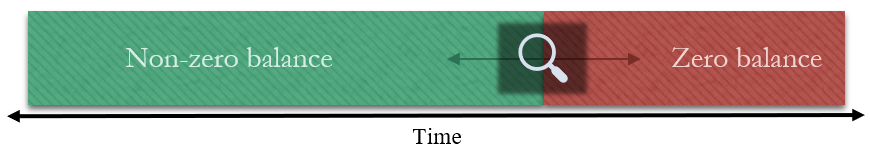}
\caption{Illustrating the `healthy' non-TZB period (green-shaded) and the `unhealthy' TZB-period (red-shaded) over the lifetime of an affected loan. The exact cut-off point between either period type is uncertain, as represented by the magnifying glass.}\label{fig:ProblemDesc}
\end{figure}

A prominent example of such a data error is the potential for loan accounts to exhibit excess (or false) repayment history beyond the true but latent closure date. These false observations appear as a series of zero-valued (or exceedingly small) month-end balances that trail the end of observed repayment histories. We shall call such cases \textit{trailing zero-valued balances} (TZB). These TZB-cases arise due to historical issues when migrating data between disparate computer systems, as well as operational failures in managing loan accounts that ultimately compromise their timely closure.
The actual (but latent) closure dates of some loan accounts can therefore precede the last dates observed from data, thereby resulting in excess monthly records (or `false' history) between these two dates; see \autoref{fig:ProblemDesc}. Naturally, it is non-trivial even to detect TZB-cases since no single definition exists by which "small balances" can be isolated; it could be ZAR 1.00 or ZAR 10,000.
As for implications, the excess history of a TZB-case renders the observed endpoint thereof as ambiguous and inconsistent. This ambiguity, as explained by \citet{schober2018survival}, will bias any subsequent survival estimates of the true but latent time-to-write-off, thereby inviting model risk and compromising the survival model itself.
Moreover, the realised LGD-value becomes upwardly biased since the resolution period over which cash flows are discounted is too long. While a larger LGD-value may prudently reflect operational failures within the bank, the resulting bias disagrees with the true credit risk profile of the affected loan. As such, the subsequent ECL-estimate is higher than strictly necessary, and this premium poses as an opportunity cost for ignoring these erroneous TZB-cases.

The obvious remedy is to remove the TZB-affected part from a loan's history, i.e.,  move the observed but false endpoint earlier, at least to the first instance of a zero-valued (or suitably small) balance. The excess history beyond this true but latent endpoint can then be discarded, not unlike a surgeon excising a malignant tumour.
However, manually finding this latent endpoint for each affected loan would quickly become challenging, especially given the typically large credit datasets in retail lending; often containing hundreds of thousands of loans over multiple decades. 
Moreover, these diminutive balances that trail the ends of credit histories can vary across accounts and can therefore defy the definition of the endpoint itself. In particular, isolating such a trail will naturally depend on the exact definition of a "small or immaterial balance", beyond which the remaining history is deemed as practically zero-valued and therefore best deleted. This small-balance definition will itself inform the overall prevalence of TZB-cases, and quite significantly so, as we will demonstrate later.
Small-balance definitions that are too small will incorrectly sequester only a portion of account lifetimes into supposed TZB-histories, which may very well be too short; thereby leaving behind `cancerous' stumps of data whilst failing to solve the problem.
Conversely, small-balance definitions that are too large will excessively elongate TZB-histories and inflate their prevalence; ultimately cutting away into otherwise credible data (not unlike an overzealous surgeon), thereby compromising the subsequent modelling.
Therefore, the question becomes this: how small exactly is a "small balance" in finding the true endpoint, though without discarding `healthy' history?
This question clearly suggests an optimisation framework for finding the starting point of the TZB-period, such that this true endpoint is neither too early nor too late. 
Optimisation itself, as explained in \citet{boyd2004convex}, is the process of finding the best solution to a quantitative problem whilst meeting certain constraints. In our context, the number of such possible solutions is infinite since the size of a small-balance definition can theoretically assume any non-negative currency amount. Evaluating any solution therefore requires an appropriate objective function $f$ that can simultaneously reward or penalise various aspects of such a solution in finding the true endpoint.

Accordingly, we contribute both such an $f$ and a broader optimisation framework that systematically evaluates the costs/benefits of any small-balance definition in isolating the latent TZB-period. This novel framework is called the \textit{TruEnd-procedure}, which finds the \textit{best} small-balance definition or policy, denoted as $b^*\in\mathbb{R}_{\geq0}$, for a given loan portfolio. Any latent TZB-cases can then be identified and duly discarded using this $b^*$-policy, thereby correcting this data error. Doing so would intuitively improve the predicted timing of risk events (such as write-off or settlement) when using this corrected data in training survival models.
Secondly, we demonstrate the TruEnd-procedure using two datasets from the South African credit market: residential mortgage accounts (secured) and credit card accounts (unsecured). Our results show that the latent endpoints of accounts can be successfully isolated by adopting a uniform $b^*$-policy per dataset, as informed by the TruEnd-procedure. 
These uniform $b^*$-policies are further validated within a broader Monte Carlo setup by repeatedly subsampling from a dataset, and applying the TruEnd-procedure onto each subset.
We further demonstrate the effect of applying our procedure (vs its absence) by developing two competing survival models for the time-to-write-off.
Thirdly, and to facilitate adoption, we provide an open-source R-based implementation of the TruEnd-procedure, with full documentation and 19 annotated scripts; see \citet{botha2024sourcecode}.
Ultimately, we believe our work demonstrates the sound pursuit of innovation in modelling, thereby alleviating model risk and improving the accuracy of ECL-estimates under IFRS 9.

This paper is structured as follows. In \autoref{sec:method}, we present the TruEnd-procedure and motivate its design using a real-world case study.
We calibrate and demonstrate this procedure in \autoref{sec:results_optimisation} using real-world data. This demonstration culminates in selecting the best $b^*$-policy from the ensuing optimisation results.
In \autoref{sec:results_impactStudy}, we investigate the impact of imposing this particular $b^*$-policy on the same data, thereby discarding the resulting TZB-periods across all duly affected accounts. In this impact study, the aforementioned survival models are compared and discussed. Finally, we conclude the study in \autoref{sec:conclusion}.
\section{A method for finding trailing zero-valued balances (TZB)}
\label{sec:method}

Consider a portfolio of $N$ loans, indexed using $i=1,\dots,N$, where each loan account $i$ is sampled over time $t$ up to its observed (but possibly false) lifetime $T_i$, which is measured in calendar months.
By definition, account $i$ has a TZB-period of minimum length $l\geq 1$ month when the month-end balances during such a period are persistently zero or close to zero. Let $B_i\geq 0$ denote the account's currency balance (denominated in ZAR), which is a random variable that is sequentially measured at month-end times $t=1,\dots,T_i$ for $i$. This yields the column vector $\boldsymbol{B}_i = \left[B_{i1}, B_{i2}, \dots, B_{iT_i} \right]^{\text{T}}$, also represented as the time series $\left\{B_{it} \right\}_{t=1}^{T_i}$. 
Given an affected account $i$, its TZB-period has a starting point $t_{z}(i)\geq1$ whereafter most of the remaining observed balances $\left\{ B_{it} \right\}_{t\geq t_z(i)}$ are noticeably lower when compared to the preceding balances $\left\{ B_{it} \right\}_{t<t_z(i)}$ prior to the TZB-period. 

In finding this $t_{z}(i)$-point for an affected account $i$, one can examine the elements of $\boldsymbol{B}_i$; itself effectively becoming a control variable within a broader optimisation context. 
More formally, let $b\geq0$ be a constant (or threshold) respective to the domain of $ \boldsymbol{B}_i$. 
Then, a TZB-account is identified whenever it has a period $t=t_z(i),\dots,T_i$ such that $B_{it}\leq b$ at every point $t$ during this period. The "true end" of such a TZB-account therefore occurs at time $t_z(i)-1$ whereat the last small balance can credibly occur, which is also $\leq b$ by definition.
As such, let $\mathcal{S}_\mathrm{T}$ be the subset of all TZB-accounts within a portfolio, which is defined for a particular $b$-value and the \textit{control matrix} $\boldsymbol{B}=\left[ \boldsymbol{B}_1, \dots, \boldsymbol{B}_N \right]$ as 
\begin{equation} \label{eq:set_tzb}
    \mathcal{S}_{T}\left(\boldsymbol{B},b \right) = \big\{ i \ \big\vert \ \exists \ t'\in[1,T_i] \ : B_{it} \leq b \quad \text{for} \ t=t',\dots,T_i \quad \text{s.t} \quad T_i - t' \geq l \big\} \, .
\end{equation}
Accordingly, the $t_z(i)$-point is the earliest moment $(t'+1)$ of such a TZB-period, barring the last small balance that can credibly occur at $t'$ (hence the `$+1$'-term), and is given by the procedure
\begin{equation} \label{eq:t_z}
    T_z(\boldsymbol{B}_i, b): \quad t_z(i) = ( t' + 1) : B_{it} \leq b \quad \text{for} \ t=t',\dots,T_i \quad \text{s.t} \quad T_i - t' \geq l  \quad \forall \ i\in\mathcal{S}_\mathrm{T}\left(\boldsymbol{B},b \right) \, .
\end{equation}
For illustration purposes, let $Z_{it}\left(B_{it}, b\right) \in \{0,1\}$ be a Boolean-valued decision function that indicates an element's membership to a TZB-period at $t$, defined for any loan $i$ as 
\begin{equation} \label{eq:TZB_Decide}
    Z_{it}\left(B_{it}, b\right) = 
    \begin{cases}
    1 \quad \text{if} \ B_{it} \leq b \quad \text{for} \ t=T_z(\boldsymbol{b}_i, b),\dots,T_i \\
    0 \quad \text{otherwise}
    \end{cases} \, .
\end{equation}

\begin{table}[ht!]
\caption{Finding the TZB-start point $t_z$ for a single real-world account $i$ (that was settled early) by evaluating $T_z(\boldsymbol{B}_i,b)$ from \autoref{eq:t_z} against $b=500$ using $\boldsymbol{B}_i$ as control. Membership to a TZB-period is indicated over time $t$ using $Z_t(B_{it},b)$ from \autoref{eq:TZB_Decide}.
The green-shaded row is the account's true end at time $t_z(i)-1$, while the orange-shaded rows constitute the TZB-period, which is marked for deletion. Given $t_z=62$, the measures $M_1\left(t_z\right)$ and $M_2\left(t_z,6\right)$ from \crefrange{eq:Measure1}{eq:Measure2} are estimated accordingly.}
\label{tab:Finding_TZB_start}
\centering
\begin{tabular}{p{1.8cm} p{1.5cm} p{1.5cm} p{1.8cm} p{1.8cm} p{0.4cm} p{0.3cm}}
\toprule
\textbf{Date} & \textbf{Time} $t$ & \textbf{Principal} & \textbf{Balance} $B_t$ & \textbf{Membership} $Z_t(B_t, b)$ \\ \midrule
\multicolumn{7}{c}{...} \\
\small{2011-08-31} & 56 ($t_z-7$)  & \small{R 90,000} & \small{R 6,028.16} & 0 & \multirow{6}{*}{}\rdelim\}{6}{1cm}[{\small{\rotatebox[origin=c]{90}{Non-TZB period}}}] & \multirow{6}{*}{\rot{\footnotesize{$M_2(t_z,6)=6,053.66$}}} \\
\small{2011-09-30} & 57 & \small{R 90,000} & \small{R 6,078.47} & 0  \\
\small{2011-10-31} & 58 & \small{R 90,000} & \small{R 7,954.24} & 0 \\
\small{2011-11-30} & 59 & \small{R 90,000} & \small{R 8,018.80} & 0 \\
\small{2011-12-31} & 60 & \small{R 90,000} & \small{R 8,085.82} & 0 \\ 
\rowcolor[HTML]{B5FFB5}  
\small{2012-01-31} & 61 ($t_z-1$) & \small{R 90,000} & \small{R 156.47} & 0 \\
\rowcolor[HTML]{FFE1BD}  
\small{2012-02-29} & 62 ($t_z$) & \small{R 90,000} & \small{R 163.30} & 1 & \cellcolor{white} \multirow{8}{*}{}\rdelim\}{8}{1cm}[{\small{\rotatebox[origin=c]{90}{TZB-period}}}] & \cellcolor{white} \multirow{8}{*}{\rot{{$M_1(t_z)=161.36$}}} \\
\rowcolor[HTML]{FFE1BD}
\small{2012-03-31} & 63 & \small{R 90,000} & \small{R 170.28} & 1 \\
\rowcolor[HTML]{FFE1BD}
\small{2012-04-30} & 64 & \small{R 90,000} & \small{R 177.26} & 1 \\
\rowcolor[HTML]{FFE1BD}
\small{2012-05-31} & 65 & \small{R 90,000} & \small{R 184.33} & 1 \\
\rowcolor[HTML]{FFE1BD}
\small{2012-06-30} & 66 & \small{R 90,000} & \small{R 191.42} & 1 \\
\rowcolor[HTML]{FFE1BD}
\small{2012-07-31} & 67 & \small{R 90,000} & \small{R 198.57} & 1 \\
\rowcolor[HTML]{FFE1BD}
\small{2012-08-31} & 68 & \small{R 90,000} & \small{R 205.73} & 1 \\
\rowcolor[HTML]{FFE1BD}
\small{2012-09-30} & 69 ($T_i$) & \small{R 90,000} & \small{R 0.00} & 1 \\ \bottomrule
\end{tabular}
\end{table}

As an illustration, consider the credit history of a single real-world mortgage account $i$ in \autoref{tab:Finding_TZB_start}, using the threshold $b$ respective to the domain of $\boldsymbol{B}_i$. Evidently, the loan balance decreased substantially to $B_{it}=156.47$ at the (manually inferred) time of early settlement $t=61$, at which point the credit history should have logically stopped but did not. Using $b=500$ as an example small-balance definition, the TZB-indicator $Z_{it}(B_{it},b)$ then flags the entire TZB-period as 1, starting at $t_z=62$. By implication, the account's "true end" is at $t=61$. Moreover, consider the mean balance across the preceding 6-month non-TZB period $t=56,\dots,61$ and compare it against that of the supposed TZB-period thereafter. The substantial difference in the respective means (6,053.66 vs. 161.36) corroborates an otherwise clear split in the credit balances, as manifested when implicitly choosing $t_z$ by explicitly choosing $b$.

A manual evaluation of individual $t_z(i)$-points across many affected loans $i \in \mathcal{S}_{T}\left(\boldsymbol{B},b \right)$ will quickly become cumbersome without some automation. However, we can formalise the use of the arithmetic mean as a tool for evaluating any particular $t_z(i)$-value of loan $i$; and indeed, for evaluating a chosen $b$-value as the portfolio-wide TZB-threshold. This idea exploits the mean's known weakness to outliers, particularly those $B_{it}$-values close to zero within a supposed TZB-period. As such, consider the measure 
\begin{equation} \label{eq:Measure1}
    M_1\big(t_z \big) = \frac{1}{T_i-t_z} \sum_t{B_{it}} \quad \text{for} \ t=t_z,\dots,T_i \quad \text{s.t.} \quad T_i-t_z\geq l \,,
\end{equation} 
where $t_z$ is a candidate starting point of the TZB-period of a particular account. This $M_1$-measure is evaluated across the TZB-period given some $t_z$-value; i.e., $M_1$ is the affected account's mean TZB-balance. Now consider the counterweight measure 
\begin{equation} \label{eq:Measure2}
 M_2\big(t_z,\tau \big) = \frac{1}{\tau} \sum_t{B_{it}} \quad \text{for} \ t=\big( t_z-\tau-1 \big),\dots, \big( t_z-1 \big) \,,
\end{equation} 
where $M_2$ is estimated across the period (of specifiable length $\tau\geq1$) that immediately precedes the TZB-period; i.e., $M_2$ is the affected account's non-TZB mean balance. For non-TZB accounts, $M_1$ is undefined while $M_2$ simply becomes the mean balance of the last $\tau$ periods, i.e., $t_z$ becomes $T_i$. 
In the interest of expediency, we simply set the length $\tau=6$ months when calculating $M_2$; though this aspect can certainly be investigated in future\footnote{Upon the request of a reviewer, we repeated the present work after doubling this parameter to $\tau=12$ months. However, none of the results changed in any meaningful way, thereby suggesting that the TruEnd-procedure is not overly sensitive to this particular parameter, at least not for the mortgage dataset and for $\tau\in[6,12]$. Whilst these results for $\tau=12$ are not reported in this paper, the curious reader is referred to our source code in \citet{botha2024sourcecode}.} work.
More importantly, and since both measures $M_1$ and $M_2$ work in concert when varying $t_z$, the influence of either measure on the other can be tracked as follows.
Given a $t_z(i)$-value for a TZB-account $i$, let $\phi_i\big( t_z(i) \big)$ denote the degree to which $M_1\big(t_z(i) \big)$ is `contaminated' by $M_2\big(t_z(i),\tau \big)$, defined simply as \begin{equation} \label{eq:contaminationDegree}
    \phi_i\big( t_z(i) \big) = \frac{M_1\big(t_z(i) \big)}{M_1\big(t_z(i) \big) + M_2\big(t_z(i),\tau \big) } \quad \forall \ i\in\mathcal{S}_\mathrm{T}\left(\boldsymbol{B},b \right) \, .
\end{equation}

\begin{figure}[ht!]
\centering\includegraphics[width=0.75\linewidth,height=0.45\textheight]{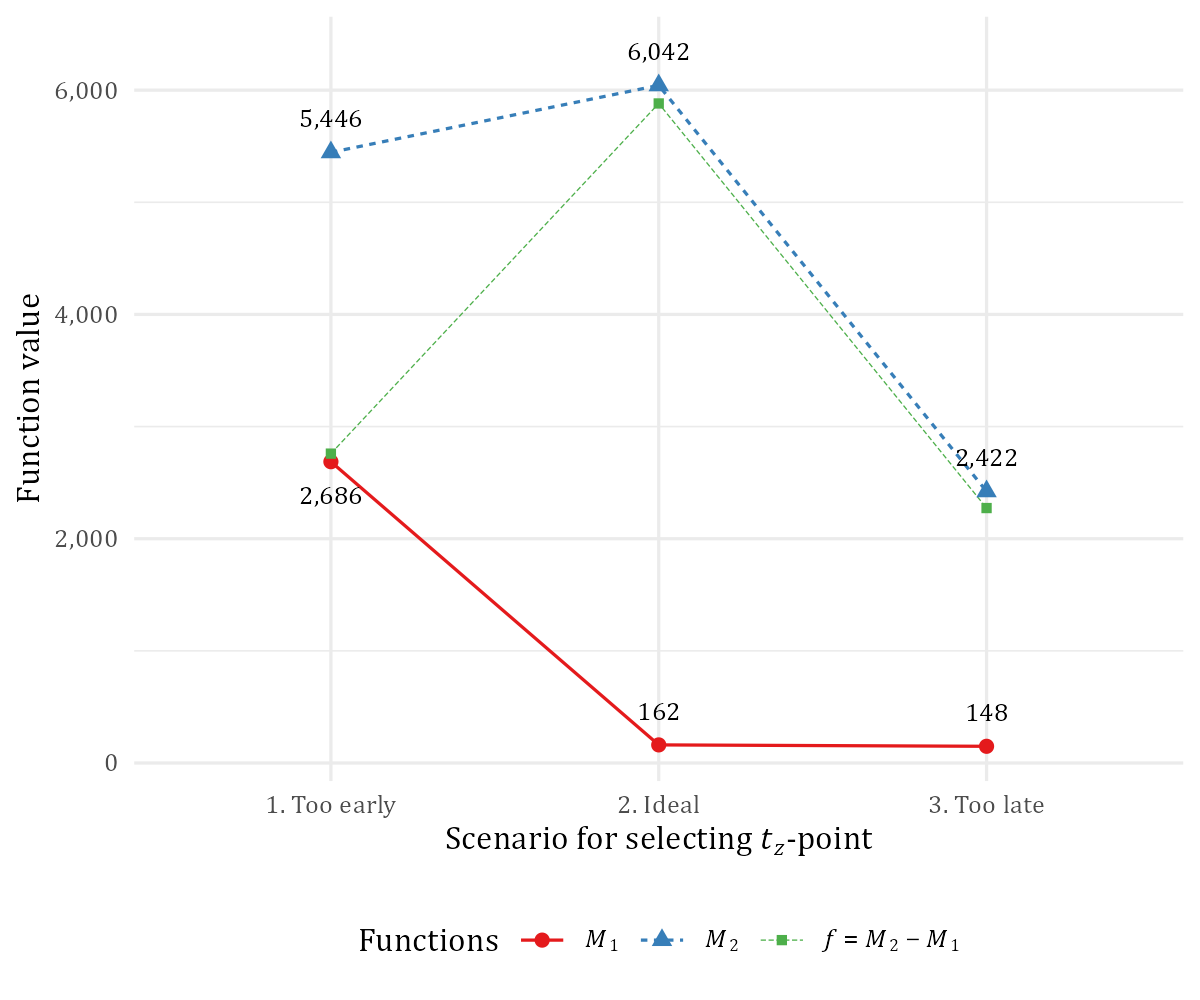}
\caption{Illustrating three independent scenarios for $M_1$ and $M_2$ from \crefrange{eq:Measure1}{eq:Measure2} in choosing the TZB-period's starting point $t_z(i)$: too early, ideal, and too late. The function $f=M_2\big(t_z(i),\tau \big) - M_1\big(t_z(i) \big)$ illustrates the combined effect across all scenarios, shown in green.}\label{fig:3Scenarios_M1M2}
\end{figure}

In principle, the $M_1\big(t_z(i) \big)$-value should be relatively closer to zero than its sibling $M_2\big(t_z(i),\tau \big)$-value, as demonstrated in \autoref{tab:Finding_TZB_start} with $t_z(i)=62$. Exceedingly greater $t_z(i)$-values (or later periods) will produce lower $M_2\big(t_z(i),\tau \big)$-values given the increasing influence of overly small $B_{it}$-values that should have been rightfully discarded. In this case, we have failed to isolate the true starting point of TZB-history. On the other hand, excessively smaller $t_z(i)$-values (or earlier periods) will wrongfully discard larger elements within $\boldsymbol{B}_i$ that denote otherwise credible history, thereby causing larger $M_1\big(t_z(i) \big)$-values. 
As an illustration, consider \autoref{fig:3Scenarios_M1M2} that compares three hypothetical scenarios for $M_1$ and $M_2$ when choosing specific $t_z(i)$-values, having used the same real-world account $i$ from  \autoref{tab:Finding_TZB_start}. 
In the ideal scenario, i.e., $t_z(i)=62$, the preceding non-TZB period is perfectly separated from the subsequent TZB-period in that the latter contains noticeably smaller balances while the former does not.
The "too early" scenario, i.e., $t_z(i)=56$, shows the increasing contamination in $M_1$ from including the obviously larger $B_{it}$-values within the TZB-period, specifically those six extra elements at $t=56,\dots,61$ in \autoref{tab:Finding_TZB_start}. Accordingly, the $M_1$-value increases substantially from its previous value under the ideal scenario.
Conversely, the "too late" scenario, i.e., $t_z(i)=66$, demonstrates the debilitating effect on the $M_2$-value (itself decreasing substantially) when including 
minuscule $B_{it}$-values within the non-TZB period. This effect demonstrates the weakness of the mean to these outlying small balances.

Put differently, a delayed $t_z$-point will yield smaller $M_2(t_z,\tau)$-values whereas a premature $t_z$-point will produce larger $M_1(t_z)$-values. Both factors suggest that the ideal $t_z(i)$-point for a TZB-account $i$ can be found by maximising $M_2$ and minimising $M_1$ with respect to $t_z(i)$, which is itself controlled by varying $b$ in $T_z(\boldsymbol{B}_i, b)$ from \autoref{eq:t_z}. Doing so implies maximising the loan-level objective function
\begin{equation} \label{eq:Objective_loan}
    l_i\big( t_z(i) \big) = w \cdot M_2\big(t_z(i),\tau \big) - M_1\big(t_z(i) \big) \quad \text{for} \ w\in[0,1] \quad \text{and} \quad\forall \ i\in\mathcal{S}_\mathrm{T}\left(\boldsymbol{B},b \right)\,.
\end{equation}
The specifiable weight $w$ in \autoref{eq:Objective_loan} is simply a mathematical expedient, which is motivated as follows.
In isolating the TZB-account's "true end", any reasonable $t_z(i)$-value also implies that the associated $M_2\big(t_z(i),\tau \big)$-value will most likely be orders of magnitude larger than its sibling $M_1\big(t_z(i) \big)$-value. In turn, this discrepancy in domain sizes imply that the value of $M_2$ will remain relatively unchanged when subtracting that of $M_1$, thereby rendering the latter as meaningless and ruining any subsequent optimisation.
Accordingly, we introduce the specifiable weight $w$ in \autoref{eq:Objective_loan} by which $M_2$ should be scaled towards matching the domain of $M_1$, even if only approximately.

Finding a single portfolio-level threshold $b$ implies evaluating \autoref{eq:Objective_loan} across many TZB-accounts. For a particular $(\boldsymbol{B},b)$-tuple (or configuration), the final portfolio-level objective function therefore becomes 
\begin{equation} \label{eq:Objective_portfolio}
    f(\boldsymbol{B}, b) = \frac{1}{\bar{s}} \sum_{i}{ l_i\big ( T_z(\boldsymbol{B}_i, b) \big) } \quad \text{for} \ \forall \ i \in \mathcal{S}_\mathrm{T}(\boldsymbol{B}, b) \, .
\end{equation}
Since the size and composition of $\mathcal{S}_\mathrm{T}(\boldsymbol{B},b)$ will change in tandem with each $(\boldsymbol{B},b)$-configuration, the eventual optimisation will undoubtedly be affected by sampling variability. In mitigating this effect, the summation in \autoref{eq:Objective_portfolio} is therefore weighted by the sample standard deviation $\bar{s}$ of the loan-level $l_i\big ( T_z(\boldsymbol{B}_i, b) \big)$-values within the TZB-set $\mathcal{S}_\mathrm{T}(\boldsymbol{B},b)$. More formally, let $n_s$ denote the size (or cardinality) of $\mathcal{S}_\mathrm{T}(\boldsymbol{B}, b)$, whereafter we estimate the mean $\bar{l}$ of these $l_i\big ( T_z(\boldsymbol{B}_i, b) \big)$-values as 
\begin{equation} \label{eq:loss_smpMean}
    \bar{l} = \frac{1}{n_s} \sum_{i}{l_i\big ( T_z(\boldsymbol{B}_i, b) \big) } \quad \text{for} \ \forall \ i \in \mathcal{S}_\mathrm{T}(\boldsymbol{B}, b) \, .
\end{equation}
Using $\bar{l}$, we then estimate the standard deviation $\bar{s}$ as 
\begin{equation} \label{eq:loss_sampVar}
    \bar{s} = \sqrt{  \frac{1}{n_s - 1} \sum_{i}{ \bigg( l_i\big ( T_z(\boldsymbol{B}_i, b) \big) - \bar{l} \bigg)^2 } } \quad \text{for} \ \forall \ i \in \mathcal{S}_\mathrm{T}(\boldsymbol{B}, b) \, .
\end{equation}

\begin{figure}[ht!]
\centering\includegraphics[width=0.8\linewidth,height=0.45\textheight]{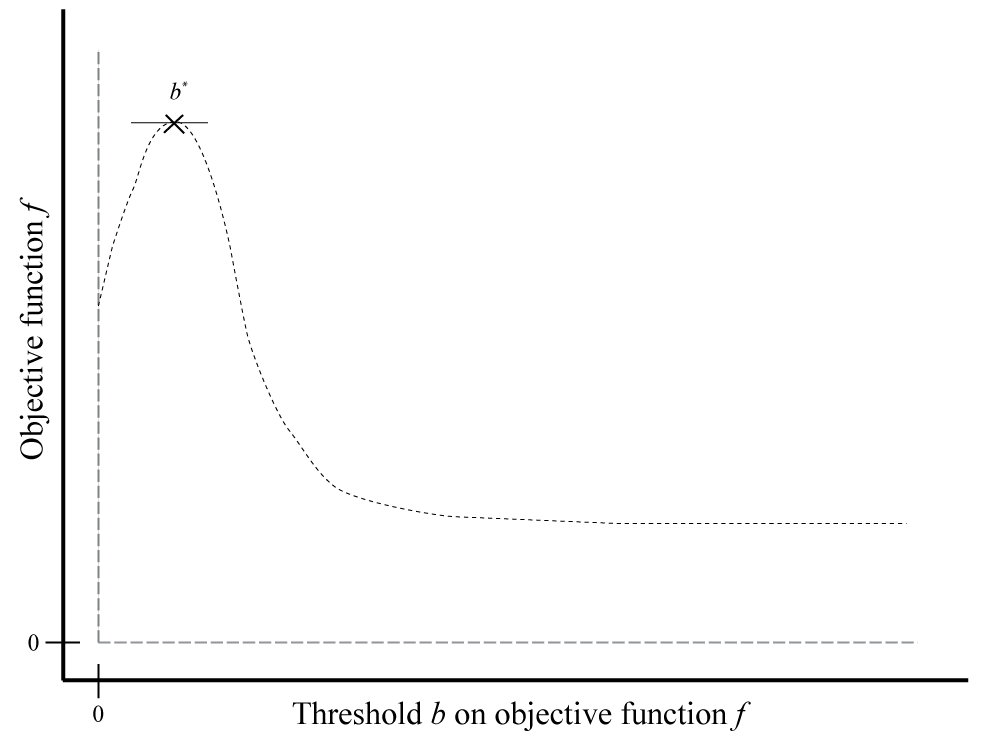}
\caption{Illustrating the TruEnd-procedure in finding the optimal threshold $b^*$ that maximises the objective function $f$ from \autoref{eq:Objective_portfolio}. Given a control variable $\boldsymbol{B}$ (e.g., balance), the elements within this times series that are smaller than $b^*$ can then be discarded, moving backwards.}\label{fig:Optimisation}
\end{figure}

Finally, the objective function $f$ from \autoref{eq:Objective_portfolio} is iteratively calculated across a range of candidate thresholds $b \in \mathcal{S}_\mathrm{b}$ within the search space $\mathcal{S}_\mathrm{b}$, having used $\boldsymbol{B}$ as a particular control variable. Similar to the LROD-procedure from \citet{botha2020LROD} and \citet{botha2020LROD_empirical}, the optimisation problem is effectively partitioned into smaller $(\boldsymbol{B},b)$-based sub-problems, where $\boldsymbol{B}$ can be any quantity used as a control and screened using $b$ as a domain-specific threshold. That said, using balances as the control makes intuitive sense in this context. Each resulting $f(\boldsymbol{B}, b)$-value is stored centrally, thereby forming a curve across all pre-chosen $b$-values for each $\boldsymbol{B}$, as demonstrated hypothetically in \autoref{fig:Optimisation}. The objective is then solved by searching for the `best' threshold $b^*$ such that $f(\boldsymbol{B}, b^*) \geq f(\boldsymbol{B}, b)$ across all candidate $b$-values within the search space $\mathcal{S}_\mathrm{b}$. This $b^*$-value is then used in finding the start times of TZB-periods $T_z(\boldsymbol{B}_i, b^*)$ of each applicable account $i$ within the underlying TZB-set $\mathcal{S}_\mathrm{T}\left(\boldsymbol{B},b^* \right)$, whereafter all monthly records are discarded for those times $T_z(\boldsymbol{B}_i, b^*),\dots, T_i$, given that these accounts actually ended at $T_z(\boldsymbol{B}_i, b^*) - 1$.
The TruEnd-procedure is therefore summarised into the following general steps:
\begin{enumerate}
    \item Enumerate the search space $\mathcal{S}_\mathrm{b}$ with thresholds $b_1,b_2, \dots, b_K$ respective to the domain of a control $\boldsymbol{B}$;
    \item Calculate $f(\boldsymbol{B}, b)$ for every $(\boldsymbol{B},b)$-configuration within $\mathcal{S}_\mathrm{b}$, and collate into the collection $\big\{ f(\boldsymbol{B},b) \big\}_{b=b_1}^{b_K}$;
    \item Search the collection $\big\{ f(\boldsymbol{B},b) \big\}_{b=b_1}^{b_K}$ for the global maximum, i.e., the optimal threshold $b^*$ is given for each $\boldsymbol{B}$ by
    \begin{equation}
        b^* =\mathop{\arg \max}_{b \, \in \, \mathcal{S}_\mathrm{b}}{\left\{ f(\boldsymbol{B}, b) \right\} } \,;
    \end{equation}
    \item Discard all records across times $t=T_z(\boldsymbol{B}_i, b^*),\dots, T_i$ for all applicable accounts $i\in\mathcal{S}_\mathrm{T}\left(\boldsymbol{B},b^* \right)$.
\end{enumerate}

\section{Illustrating the TruEnd-procedure using real-world data}
\label{sec:results_main}

The TruEnd-procedure is illustrated and tested using two rich portfolios of residential mortgages and credit card accounts, as provided by a large South African bank. These longitudinal datasets have monthly loan performance observations over time, including outstanding balances and indicators for write-off and early settlement. Mortgage accounts ($N=653,331$) were observed from January 2007 up to December 2022, whereas credit card accounts ($N=297,164$) were observed from May 2007 up to May 2023. During these periods, new accounts were continuously originated for both loan products. 
The data is extensively prepared and cleaned, as documented in the R-based codebase of \citet{botha2024sourcecode}, which includes some descriptive statistics.
Regarding the layout of this section, we present the optimisation results in \autoref{sec:results_optimisation} across both datasets. From these results, we select the optimal threshold $b^*$ that can serve as a small-balance definition, and perform a comprehensive impact study in \autoref{sec:results_impactStudy} using the mortgage dataset. This impact study is centred on comparing the dataset before and after excluding the TZB-periods, as isolated using $b^*$ as the optimal small-balance definition. 

\subsection{Optimisation results using the datasets \texorpdfstring{$\mathcal{D}_s$}{Lg} and \texorpdfstring{$\mathcal{D}_f$}{Lg}}
\label{sec:results_optimisation}

In applying the TruEnd-procedure on a loan portfolio of size $N$, a value must first be assigned to the weight $w\in[0,1]$ in \autoref{eq:Objective_loan} for down-scaling the non-TZB mean $M_2$, lest the effect of the (much smaller) TZB-mean $M_1$ becomes negligible. 
Consider that the contamination degree $\phi$ from \autoref{eq:contaminationDegree} already represents the percentage-valued contribution of $M_1$ to the sum $M_1 + M_2$; an intuitive basis from which we can derive a $w$-value at the portfolio-level. For each candidate threshold $b$, we calculate the means of both $M_1$ and $M_2$ across all applicable accounts, denoted respectively as $\bar{M}_1$ and $\bar{M}_2$. More formally, the \textit{portfolio-level} contamination degree $\bar{\phi}(\boldsymbol{B},b)$ is estimated for each $b \in \mathcal{S}_\mathrm{b}$ respective to a given control $\boldsymbol{B}$ as 
\begin{equation} \label{eq:contaminationDegree_portfolio}
    \bar{\phi}(\boldsymbol{B},b) = \frac{\bar{M}_1(\boldsymbol{B},b)}{\bar{M}_1(\boldsymbol{B},b) + \bar{M}_2(\boldsymbol{B},b)} \, ,
\end{equation}
where its components are simply estimated as
\begin{align}
     \bar{M}_1(\boldsymbol{B},b) &= \frac{1}{n_s} \sum_{i \, \in \, \mathcal{S}_T(\boldsymbol{B}, b)}{ M_1\Big ( T_z(\boldsymbol{B}_i, b) \Big) } \, , \ \text{and} \nonumber \\
    \bar{M}_2(\boldsymbol{B},b) &= \frac{1}{N} \left( \sum_{i \, \in \, \mathcal{S}_T(\boldsymbol{B}, b)}{ M_2\Big( T_z(\boldsymbol{B}_i, b), \tau \Big) } + \sum_{j \, \notin \, \mathcal{S}_T(\boldsymbol{B}, b)}{ M_2\Big( T_j, \tau \Big) } \right) \quad \text{for} \ \tau=6 \ \text{months} \, . \nonumber
\end{align}

\begin{figure}[ht!]
\centering\includegraphics[width=0.76\linewidth,height=0.45\textheight]{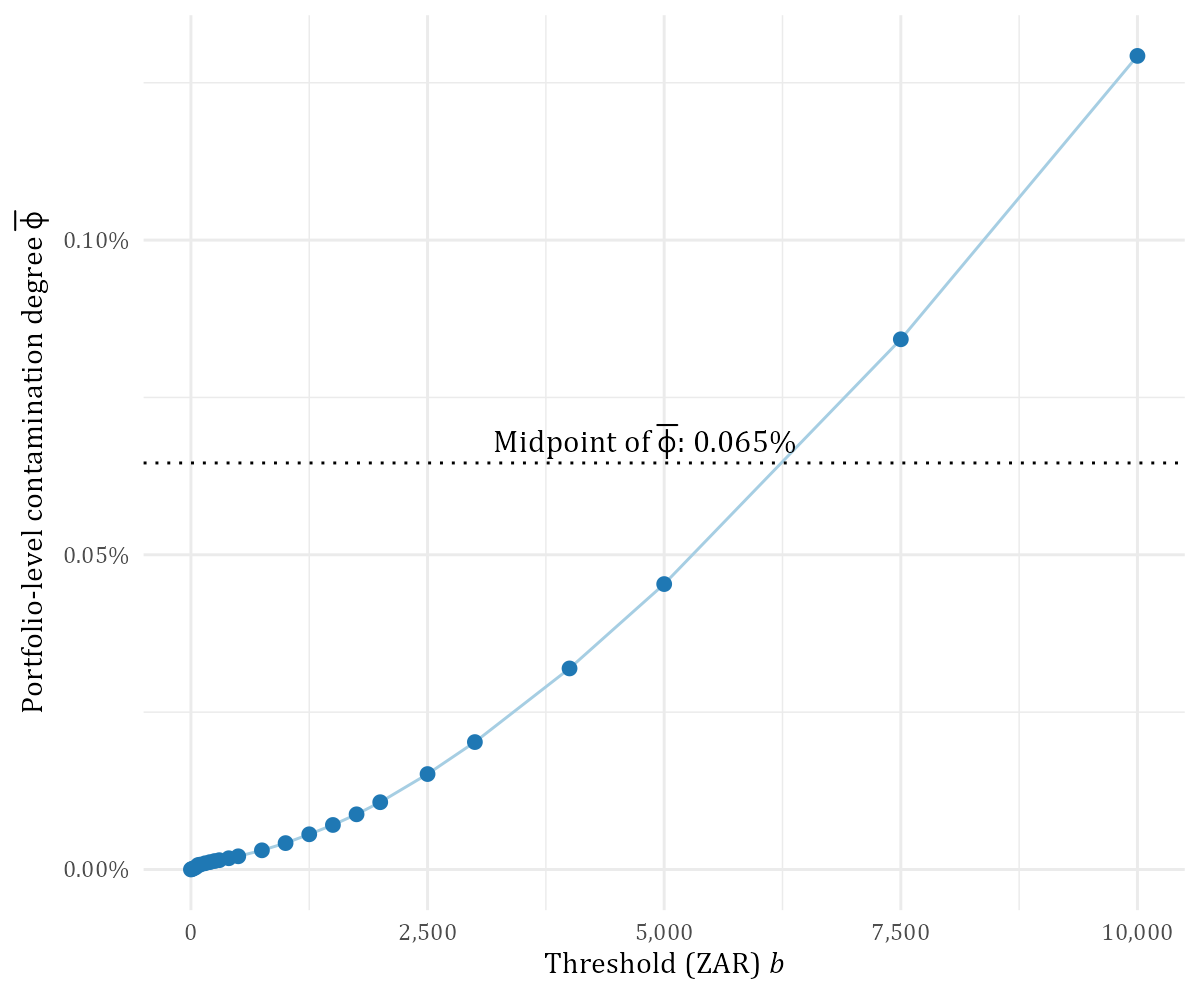}
\caption{Portfolio-level contamination degree $\bar{\phi}(\boldsymbol{B},b)$ from \autoref{eq:contaminationDegree_portfolio} across candidate thresholds $b$ respective to account balances $\boldsymbol{B}$ as control. The midpoint is overlaid, which informs the $w$-value within the TruEnd-procedure. }\label{fig:ContaminationDegrees}
\end{figure}

Using the mortgage dataset, we graph the contamination degree $\bar{\phi}(\boldsymbol{B},b)$ in \autoref{fig:ContaminationDegrees} across several candidate thresholds $b\in\mathcal{S}_\mathrm{b}$. The search space is enumerated using expert judgement\footnote{The way in which $\mathcal{S}_\mathrm{b}$ is enumerated will not only depend on the particular loan portfolio, but also on the practitioner's judgement regarding which values can intuitively constitute a "small balance" given the portfolio's context. Moreover, these particular values in $\mathcal{S}_\mathrm{b}$ were confirmed as contextually viable candidates by industry-based practitioners.} and is the same for both datasets; i.e.,  
\begin{align} \label{eq:searchspace_b}
    \mathcal{S}_\mathrm{b} = \{ & 0; 10; 25; 50; 75; 100; 150; 200; 250; 300; 400; 500; 750; 1,000; 1,250; \nonumber \\
    & 1,500; 1,750; 2,000; 2,500; 3,000; 4,000; 5,000; 7,500; 10,000 \} \,  .
\end{align}
Evidently, the $\bar{\phi}(\boldsymbol{B},b)$-quantity appears to be a monotonically increasing function of $b$, and even somewhat linear. This result is intuitively sensible since $\bar{M}_1(\boldsymbol{B},b)$ should increase in tandem with $b$, especially when denoting balances smaller than $b$ as "practically zero-valued" towards identifying TZB-periods later. Then, we define the \textit{midpoint} across all contamination degrees as the average $\bar{\phi}(\boldsymbol{B},b)$-value between the lowest and highest of such degrees; i.e., $50\%\cdot\sum{\bar{\phi}(\boldsymbol{B},b)}$ for the first and last $b$-values within $\mathcal{S}_\mathrm{b}$ from \autoref{eq:searchspace_b}.
The resulting midpoint of 0.065\% is then used as the $w$-value in \autoref{eq:Objective_loan}, thereby calibrating the TruEnd-procedure towards this particular mortgage portfolio. 
This $w$-value is similarly calibrated for the credit card portfolio, resulting in a midpoint of 3.916\%.
This rather simplistic method for finding a $w$-value produces reasonable and intuitive optimisation results, as will be shown later. However, it is imperative to acknowledge the significance of this $w$-value when calibrating the TruEnd-procedure to any particular portfolio, as our results attest. Therefore, future research can most certainly improve upon this aspect, as well as examine the impact of the chosen $\tau$-value.

\begin{figure}[ht!]
\centering
\begin{subfigure}[b]{0.49\textwidth}
    \caption{TruEnd-results from the mortgage dataset}
    \centering\includegraphics[width=1\linewidth,height=0.29\textheight]{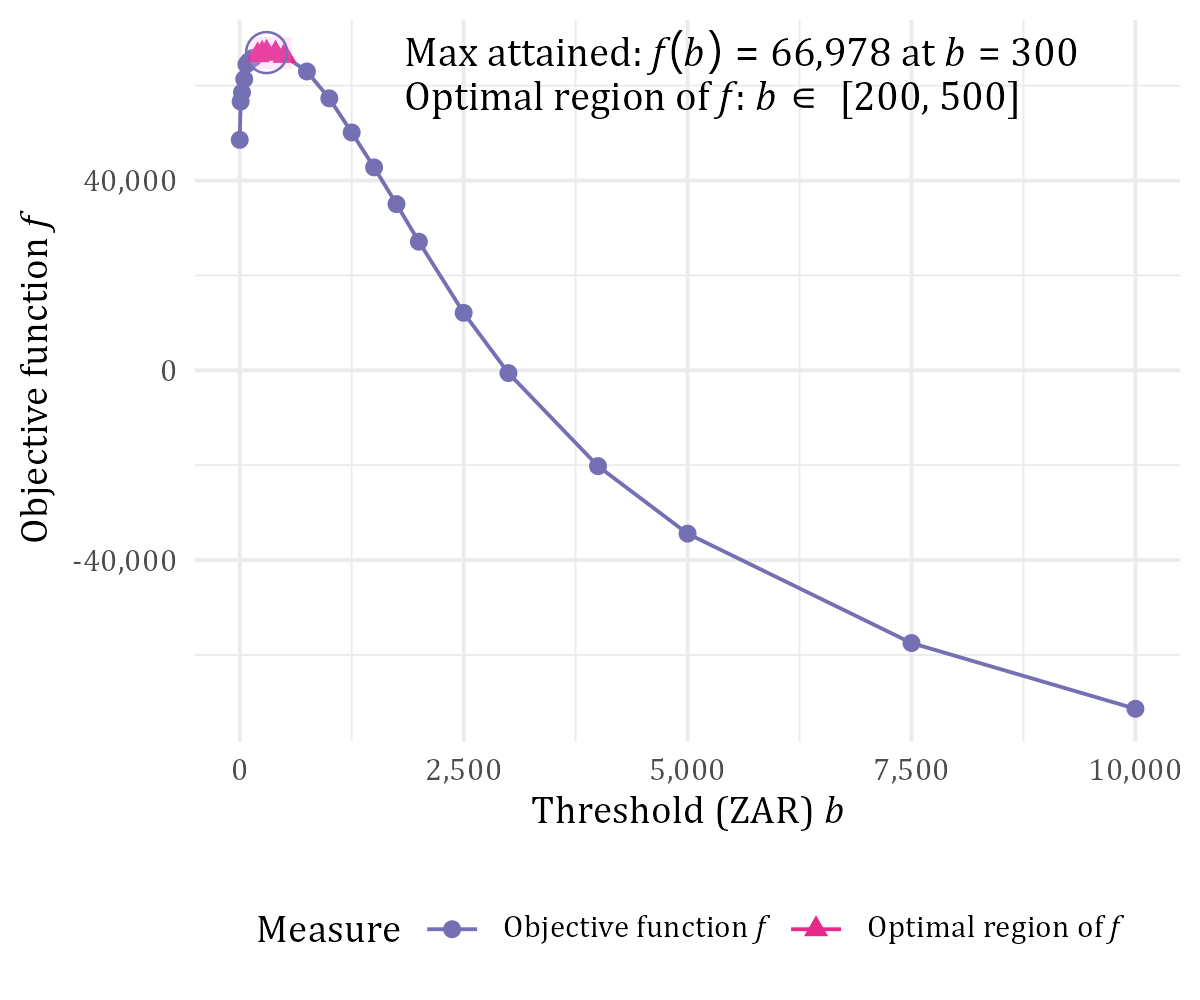}\label{fig:TruEndOptima_a}
\end{subfigure} 
\begin{subfigure}[b]{0.49\textwidth}
    \caption{TruEnd-results from the credit card dataset}
    \centering\includegraphics[width=1\linewidth,height=0.29\textheight]{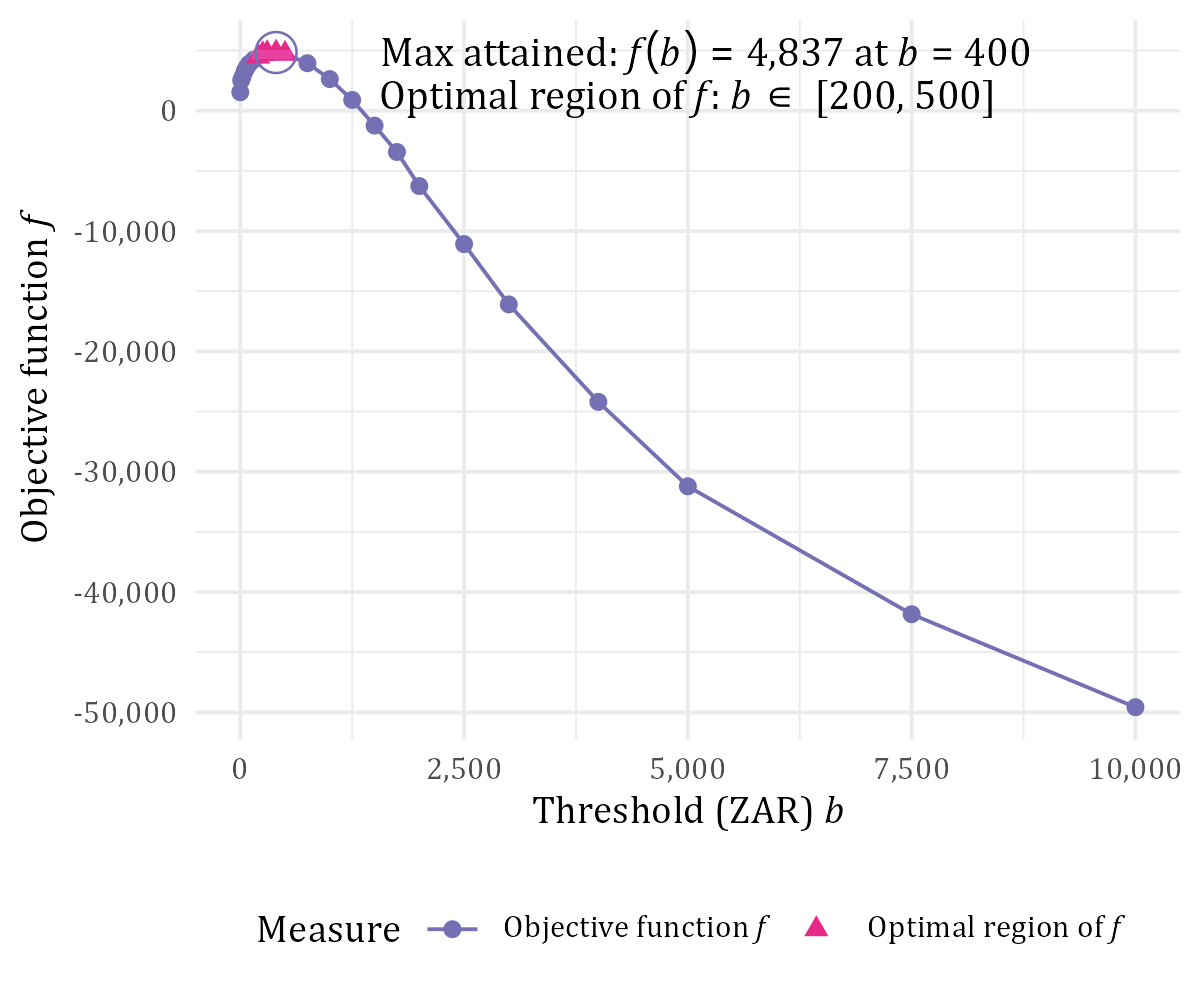}\label{fig:TruEndOptima_b}
\end{subfigure}%
\hfill%
\caption{TruEnd-results $f\left(\boldsymbol{B},b\right)$ across candidate thresholds $b$, respective to account balances $\boldsymbol{B}$ as control. The encircled point denotes the best policy $b^*$ that maximises $f$ from \autoref{eq:Objective_portfolio}. The pink triangular points that surround $b^*$ signify an optimal region in $f$ as annotated. The TruEnd-results are respectively derived in \textbf{(a)} from the mortgage dataset, and in \textbf{(b)} from the credit card dataset.}\label{fig:TruEndOptima}
\end{figure}

We present the TruEnd-results in \autoref{fig:TruEndOptima} across thresholds $b\in\mathcal{S}_\mathrm{b}$ from \autoref{eq:searchspace_b} and for the same control variable $\boldsymbol{B}$ (balances), having iteratively estimated $f(\boldsymbol{B}, b)$ from \autoref{eq:Objective_portfolio} on both of the datasets (mortgages \& credit cards). The results are strikingly similar in that both datasets produce an $f$-curve with the same shape. Moreover, the optimal $b^*$-value is ZAR 300 for the mortgage dataset, and ZAR 400 for the credit card dataset. Both $b^*$-values can serve as a small-balance definition (or policy) within their respective portfolios.
The $b$-values within the optimal region\footnote{This region is determined simply by calculating the Euclidean distance $d_E(x_i)$ between each point $x_i,i=1,\dots,K$ corresponding to thresholds on the $f$-curve, and the global maximum $x^*$. All $x_i$-points with $d_E(x_i)\leq c$ are said to be within the "optimal region", where $c$ is a cut-off that equates to a suitably-chosen lower quantile (say 20-30\%) of the resulting distribution of $d_E$-values.}, i.e., $b\in[200,500]$, do not differ significantly from one another on their face value; as corroborated by the relatively small differences in their $f(\boldsymbol{B}, b)$-values. By implication, any of these $b$-values can serve quite intuitively as small-balance definitions in practice, despite $b^*$ being the `best' such value.
Conversely, more extreme $b$-values such as $b<100$ or $b\geq2,500$ are correctly disqualified in \autoref{fig:TruEndOptima} given their increasing sub-optimality.
These results therefore provide high-level assurance on the soundness of the TruEnd-procedure and its optima.

\begin{figure}[ht!]
\centering\includegraphics[width=0.65\linewidth,height=0.40\textheight]{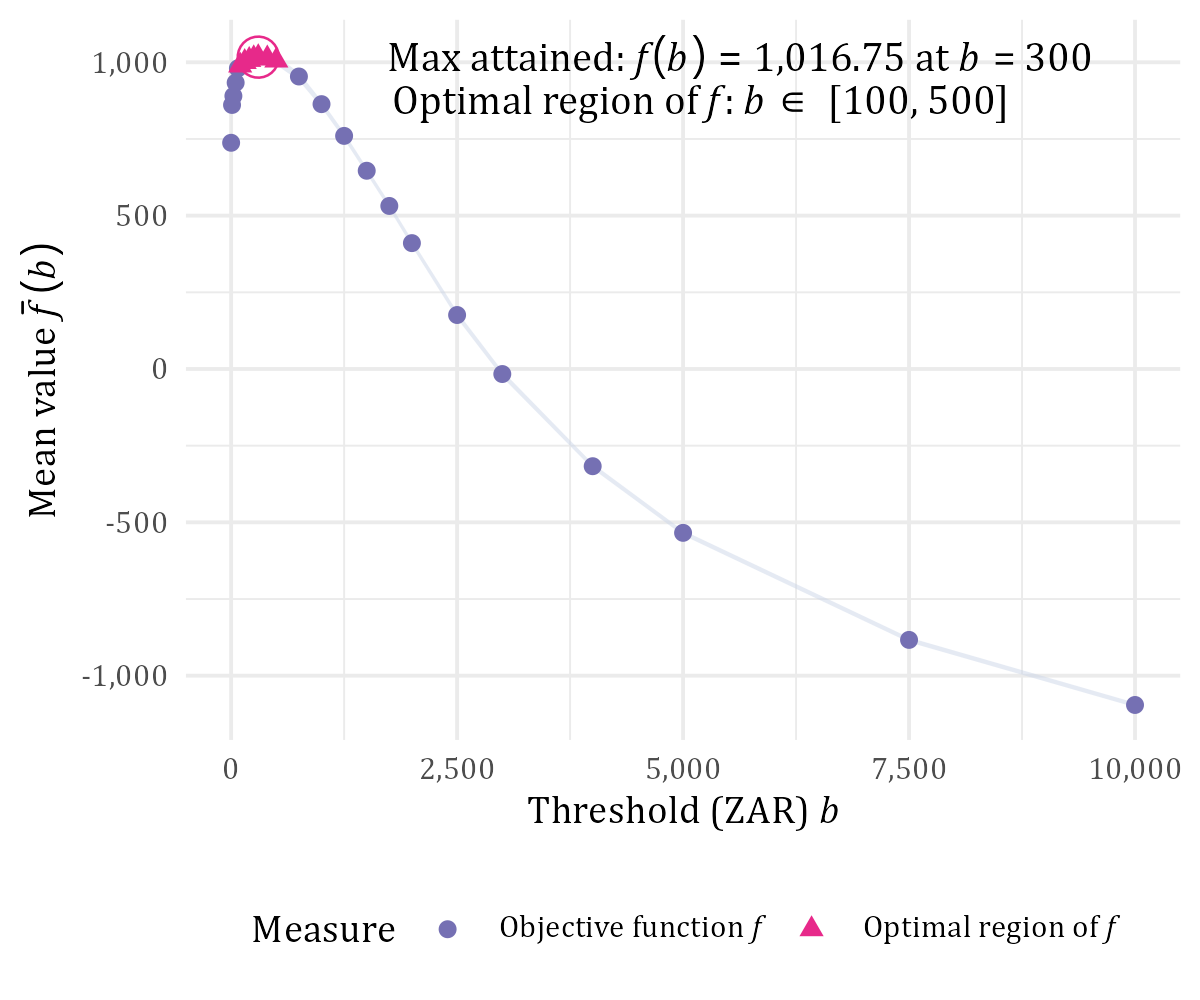}
\caption{Mean TruEnd-results $\bar{f}(b)$ across candidate thresholds $b$ within the mortgage dataset, having used the arithmetic mean within a broader Monte Carlo setup of 100 subsampling runs. 95\% confidence intervals are overlaid at each point $\bar{f}(b)$, calculated as $\bar{f}(b)\pm 1.96\left(s_b/\sqrt{100}\right)$, where $s_b$ denotes the corresponding sample standard deviation across the 100 subsampling runs. Graph design follows \autoref{fig:TruEndOptima}.}\label{fig:MonteCarlo_Optima}
\end{figure}

Given the extremely large sample sizes of our datasets, we deliberately subsample the full mortgage dataset $\mathcal{D}_f$ into a smaller set $\mathcal{D}_s$ using clustered random sampling. Doing so would allow us to investigate some statistical properties of the optima yielded by the TruEnd-procedure. In particular, we are interested in sampling the distribution of optima towards investigating the underlying uncertainty, which implies using a Monte Carlo setup. This $\mathcal{D}_s$-set contains the entire repayment histories of only 10,000 mortgage accounts that were randomly selected for a given seed value, where $\mathcal{D}_s$ effectively constitutes only 1.5\% of all accounts within $\mathcal{D}_f$. The TruEnd-procedure can then be applied on $\mathcal{D}_s$. We repeat this subsampling-process for another seed value within a broader Monte Carlo setup, and re-apply the TruEnd-procedure on the subsequent dataset. Following some experimentation, the Monte Carlo setup spans $R=100$ subsampling runs, which balances rigour against computation time. Each subsampling run $r=1,\dots,R$ yields the result set $\left\{R_{b_1r},\dots,R_{b_Kr} \right\}$ across the thresholds $b=b_1,\dots,b_K$ in $\mathcal{S}_\mathrm{b}$, where each $R_{br}=f(\boldsymbol{B},b)$ using \autoref{eq:Objective_portfolio}. These results may then be aggregated across the subsampling runs using the arithmetic mean, which is simply calculated as $\bar{f}(b)=(1/R)\sum_{r=1}^R{R_{br}}$ for a given $b$-value. We present the Monte Carlo results in \autoref{fig:MonteCarlo_Optima}, which includes a faintly visible 95\% confidence interval. The thinness of these confidence intervals attest to little uncertainty that underpins the derived optima.
Moreover, these results are quite clearly remarkably similar to those obtained in \autoref{fig:TruEndOptima_a} using the full set $\mathcal{D}_f$, and even the optimal $b^*$-value remains the same at ZAR 300. Together with the thin confidence intervals, this similarity implies that the TruEnd-procedure achieves stability in its optima, which provides assurance on the procedure's robustness against sampling-related uncertainty.

\begin{figure}[ht!]
\centering
\begin{subfigure}[b]{0.49\textwidth}
    \caption{TZB-prevalence within the mortgage dataset}
    \centering\includegraphics[width=1\linewidth,height=0.29\textheight]{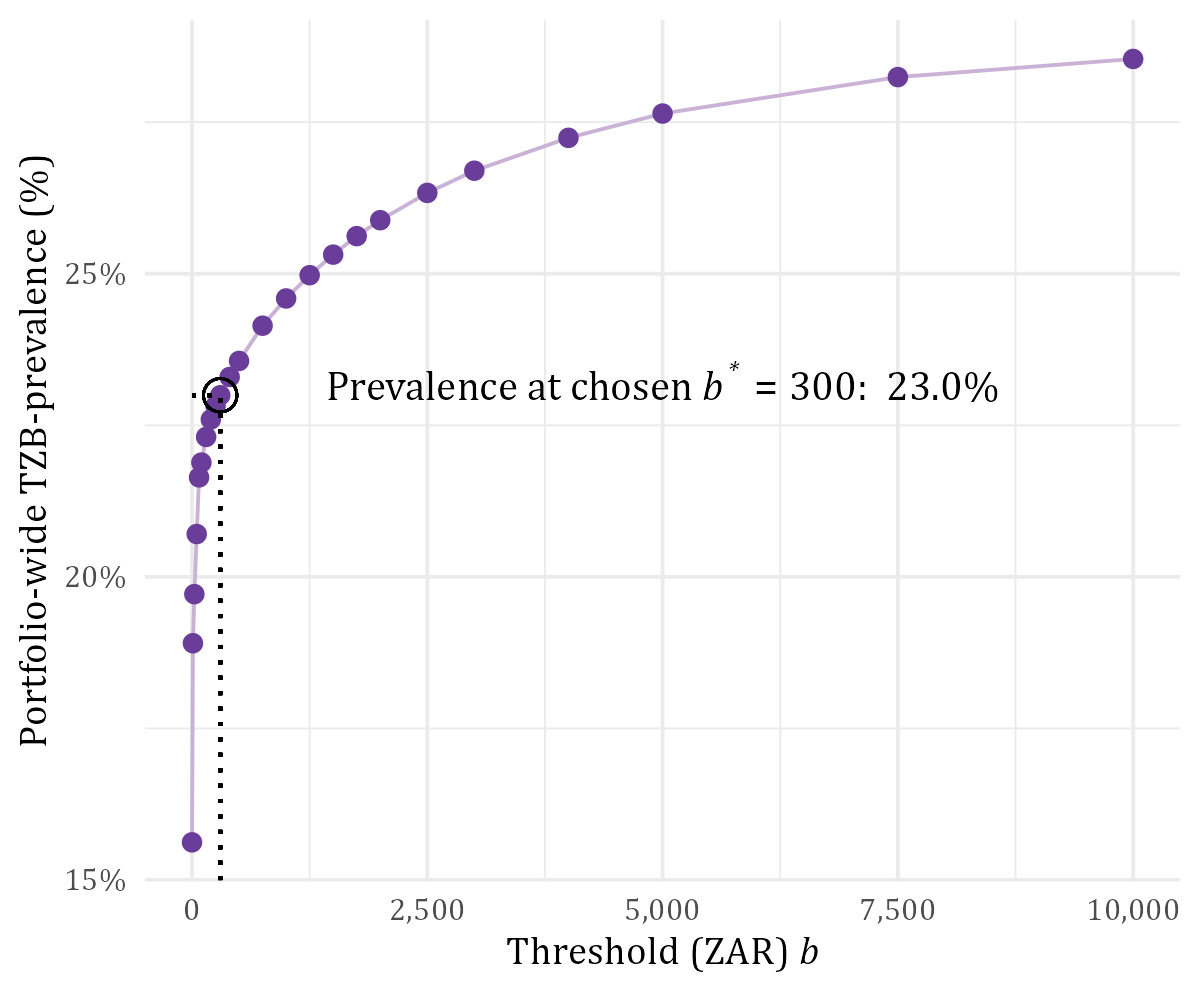}\label{fig:TZB_Prevalence_a}
\end{subfigure} 
\begin{subfigure}[b]{0.49\textwidth}
    \caption{TZB-prevalence within the credit card dataset}
    \centering\includegraphics[width=1\linewidth,height=0.29\textheight]{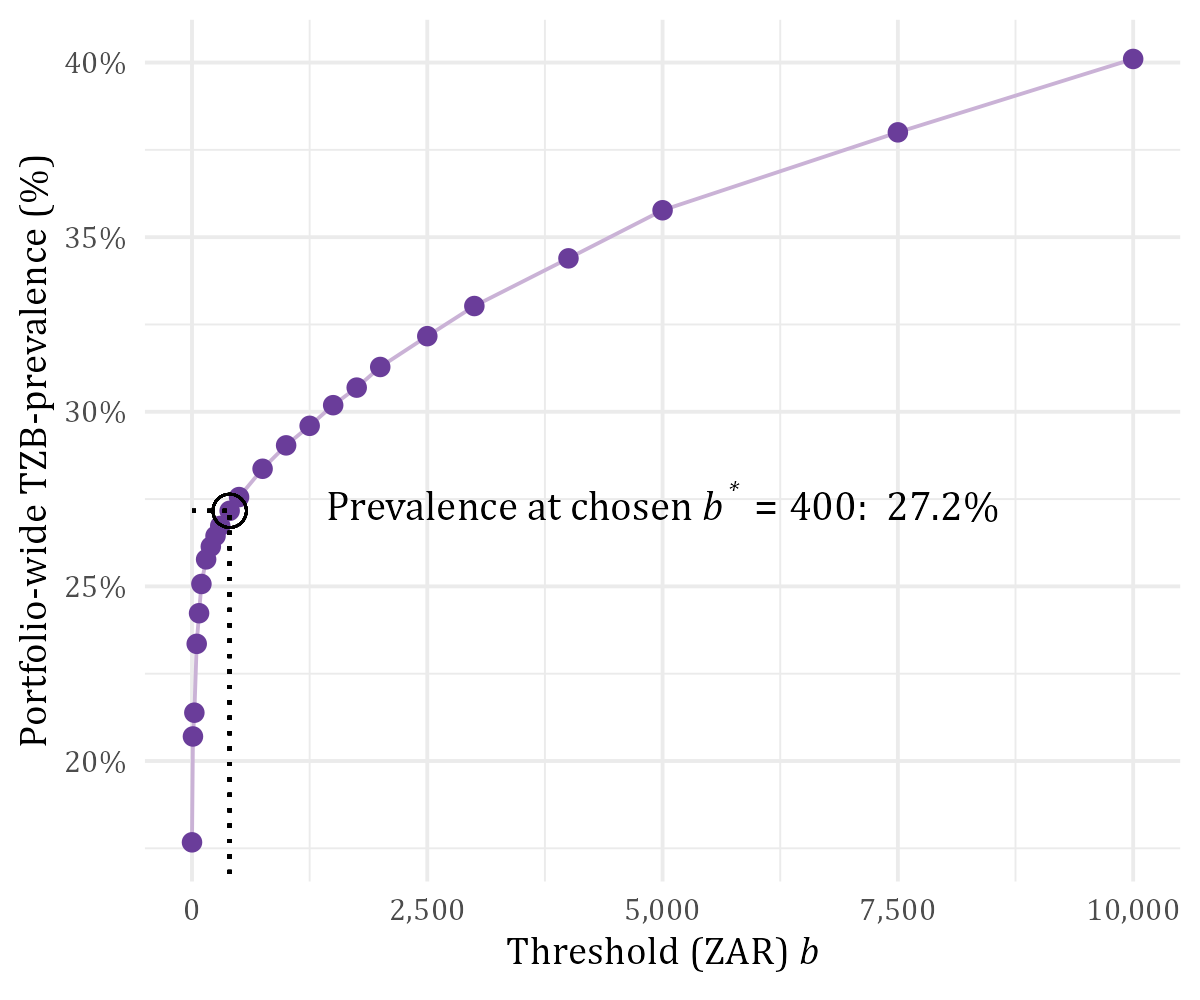}\label{fig:TZB_Prevalence_b}
\end{subfigure}%
\hfill%
\caption{The TZB-prevalence rate across candidate thresholds $b$, respective to account balances $\boldsymbol{B}$ as control within two datasets. Given the optimal $b^*$-policies from \autoref{fig:TruEndOptima}, the associated TZB-prevalence rate is overlaid. The panels \textbf{(a)-(b)} correspond respectively to the mortgage and credit card datasets.}\label{fig:TZB_Prevalence}
\end{figure}

In principle, the prevalence of any phenomenon depends on the definition thereof. Likewise, the portfolio-wide prevalence of accounts with a TZB-period will also depend on the exact definition of a "small balance" used in isolating such TZB-periods, i.e., the candidate $b$-values. 
Given a particular $b$-value, recall the decision function $Z_{it}(B_{it}, b)$ from \autoref{eq:TZB_Decide} that indicates at time $t$ whether account $i$ is in its TZB-period or not. We aggregate these 0/1-values for each $i$ by taking the maximum over $t$, whereafter these aggregates are denoted as $Z_i(\boldsymbol{B}_i,b)$ given each balance vector $\boldsymbol{B}_i$. These account-level TZB-indicators form a sample of size $N'$ from which the \textit{TZB-prevalence rate} can be estimated at the portfolio-level.
Expressed as $\sum_i{ Z_i(\boldsymbol{B}_i,b) / N' }$, we graph the resulting TZB-prevalence rate in \autoref{fig:TZB_Prevalence} across the same $b$-values for both datasets. Evidently, TZB-prevalence increases rapidly as $b$ increases, though its slope flattens progressively for points beyond the optimal $b^*$-policy. With a range of approximately 15-28\% for mortgages (and 17.5-40\% for credit cards), the overall TZB-prevalence is remarkably sensitive to the choice of $b$. This is especially true when considering that the TZB-prevalence effectively doubles when comparing the extremities. The optimal $b^*$-policy itself yields a TZB-prevalence rate of 23\% for mortgages and 27.2\% for credit cards. 
About a quarter of each portfolio's accounts are therefore afflicted with this pernicious TZB-error, which would have likely remained undetected if not for applying the TruEnd-procedure.

\begin{figure}[ht!]
\centering
\begin{subfigure}[b]{0.49\textwidth}
    \caption{Mean account ages}
    \centering\includegraphics[width=1\linewidth,height=0.45\textheight]{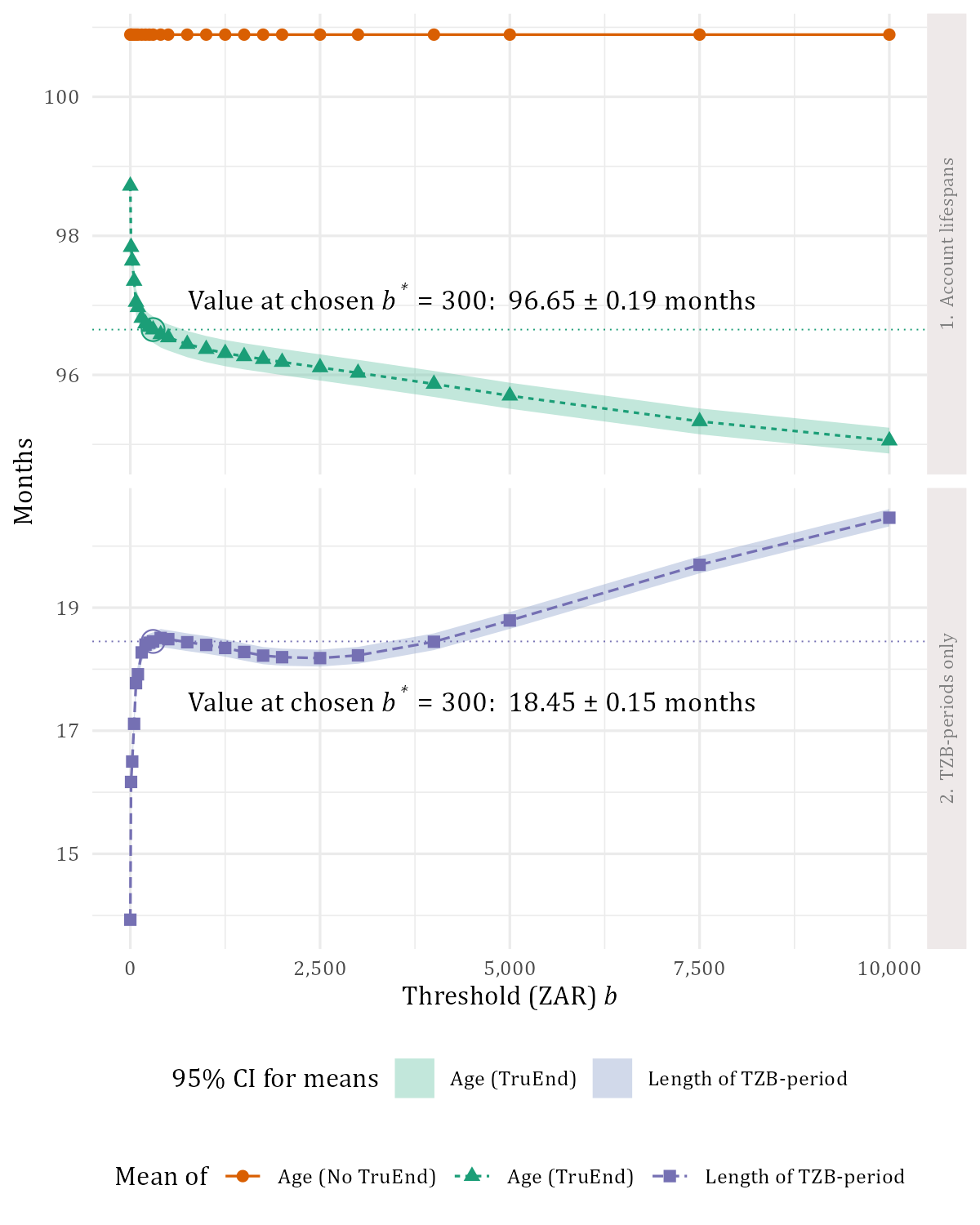}\label{fig:TruEnd_Ages}
\end{subfigure} 
\begin{subfigure}[b]{0.49\textwidth}
    \caption{Mean balances}
    \centering\includegraphics[width=1\linewidth,height=0.45\textheight]{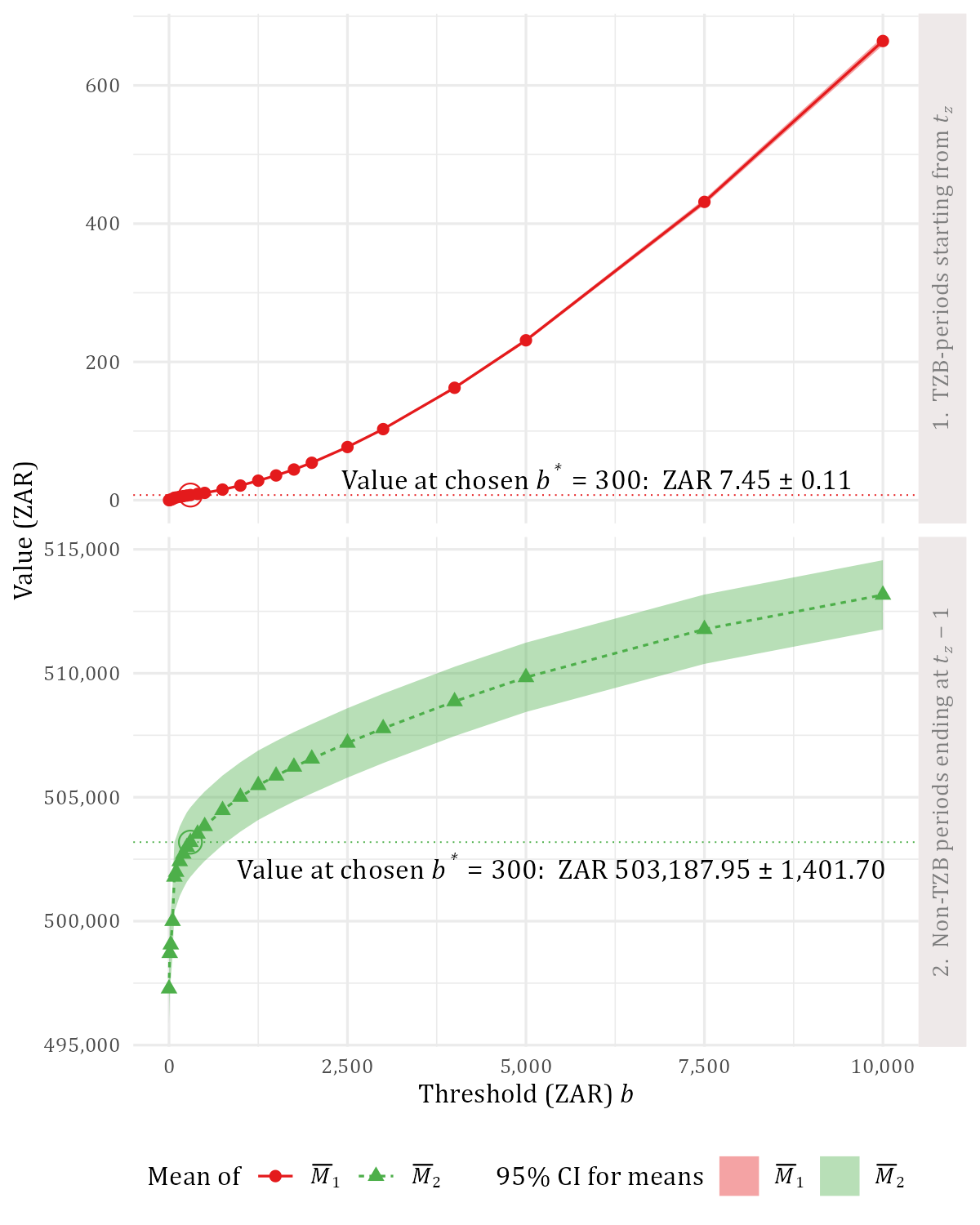}\label{fig:TruEnd_Balances}
\end{subfigure}%
\hfill%
\caption{Sample mean values $\bar{x}_b$ of various measures across candidate thresholds $b$, respective to account balances $\boldsymbol{B}$ as the control variable within the mortgage dataset. Error margins (shaded) are illustratively overlaid for each $\bar{x}_b$-estimate using 95\% confidence intervals, calculated as $\bar{x}_b\pm 1.96\left(s_b/\sqrt{n_b}\right)$, where $n_b$ denotes the sample size for a given $b$-policy, and $s_b$ denotes the corresponding sample standard deviation. Given the optimal $b^*$-policy from \autoref{fig:TruEndOptima}, the associated means are annotated. In \textbf{(a)}, the means of account ages and TZB-period lengths are shown in the respective sub-panels. In \textbf{(b)}, the portfolio-level means $\bar{M}_1(\boldsymbol{B},b)$ and $\bar{M}_2(\boldsymbol{B},b)$ from \autoref{eq:contaminationDegree_portfolio} are similarly shown.}\label{fig:TruEnd_Balances_Ages}
\end{figure}

The high TZB-prevalence rates presented thus far are certainly worrisome, given the potential for severely biased predictions from subsequent time-to-event models. 
However, the degree of such bias will inherently depend on the materiality (or impact) of TZB-cases, even if quite prevalent within a portfolio. For each candidate threshold $b$ in \autoref{fig:TruEnd_Balances_Ages}, we therefore investigate the impact of imposing such a $b$-policy, having discarded the resulting TZB-period for each affected account. In the interest of brevity, we shall restrict this investigation to the mortgage dataset, though similar results hold for the credit card dataset.
Starting at 98.7 months for $b=0$, the mean account age in \autoref{fig:TruEnd_Ages} decreases swiftly at first as $b$ increases up to the optimal $b^*$-policy, at which point the loans have a shorter average lifetime of 96.7 months. The mean length of these discarded TZB-periods also increases as $b$ grows slightly larger; i.e., from 13.9 to 18.5 months for $b=0,\dots,b^*$. The respective downward and upward trends in both means are intrinsically sensible since we fully expect the mean age to decrease when discarding TZB-periods of greater lengths as $b$ increases.
However, for larger $b$-values, the rate of change in both of these means slows down noticeably after reaching the optimal $b^*$-policy, despite the increments between successive $b$-values growing larger. These flatter slopes suggest that the TruEnd-procedure has greater impact for smaller $b$-values given the greater sensitivity to changes at those smaller locales.
Moreover, the mean loan age of 100.9 months decreases quite significantly by about 4.2 months when imposing the optimal $b^*$-policy and discarding the resulting TZB-periods; themselves having a mean length of 18.5 months and median of 5 months.
Accordingly, certain risk events such as write-off or early settlement can occur far earlier than previously thought, which has sobering implications for the degree of mistimed predictions from survival models.

Aside from analysing the age-related impact of applying the TZB-procedure, we also examine its impact in \autoref{fig:TruEnd_Balances} from the perspective of monetary value.
In particular, the portfolio-level mean $\bar{M}_1$ appears to be a monotonically increasing function of the candidate threshold $b$, which is unsurprising given the role of $b$ in qualifying balances as TZB-cases. 
Moreover, the mean balances during these TZB-periods are reassuringly small across most $b$-values, at least up to the optimal $b^*$-policy; at which point the mean is approximately ZAR 7.45. This empirical result corroborates the intuition that a TZB-period should -- by definition -- only span those last few balances that are "practically zero-valued". The monetary impact on the portfolio would therefore be minuscule when discarding these isolated TZB-periods, especially for smaller $b$-values.
Furthermore, the other portfolio-level mean $M_2$ also seems to be a monotonically increasing function of $b$. Such an increasing trend is innately sensible since small-valued outlying TZB-cases are increasingly removed from the window over which $\bar{M}_2$ is calculated, thereby `sifting' the mean.
However, and unlike $\bar{M}_1$, the rate of change in $\bar{M}_2$ is far greater for smaller values of $b\leq b^*$ than for larger values; just like for the means of loan age and TZB-period length in \autoref{fig:TruEnd_Ages}.
Assuming the optimal $b^*$-policy, we therefore conclude that TZB-cases are not only quite prevalent in affecting 23\% of mortgage accounts, but can materially prejudice the timing of risk events given the surprisingly long trail of excessive history ($\geq18$ months). Excluding these trails will correct the timing though without affecting the overall portfolio size, given the very low mean TZB-balance of $\bar{M}_1(\boldsymbol{B},b^*) = $ ZAR 7.45.
\subsection{The impact of applying the TruEnd-procedure vs its absence within the mortgage dataset}
\label{sec:results_impactStudy}

Following \autoref{sec:results_optimisation}, we duly select the optimal policy $b^*=300$ in serving as the portfolio-wide cut-off for identifying TZB-periods within the mortgage\footnote{In this impact study, we restrict our attention only to the mortgage dataset in the interest of brevity.} dataset. Put differently, $b^*$ is the `best' small-balance definition in identifying the start times of TZB-periods.
However, this $b^*$-policy is selectively applied in that the resulting TZB-periods are discarded only for terminated accounts, i.e., write-offs and settlements. Doing so is deemed prudent in limiting the potential for unwittingly overwriting the observed reality (or data) when discarding records, at least beyond what is necessary.
Furthermore, it so happens that these terminated accounts constitute 83\% of identified TZB-cases, while the remaining minority cases are still active (or right-censored) by the end of the sampling window. This rather high prevalence of terminated accounts within TZB-cases actually corroborates the original description of the problem; i.e., operational and/or system failures can delay the timely closure of accounts, thereby corrupting their observed endpoints.

\begin{figure}[ht!]
\centering
\begin{subfigure}[b]{0.49\textwidth}
    \caption{Overall account lifetimes}
    \centering\includegraphics[width=1\linewidth,height=0.3\textheight]{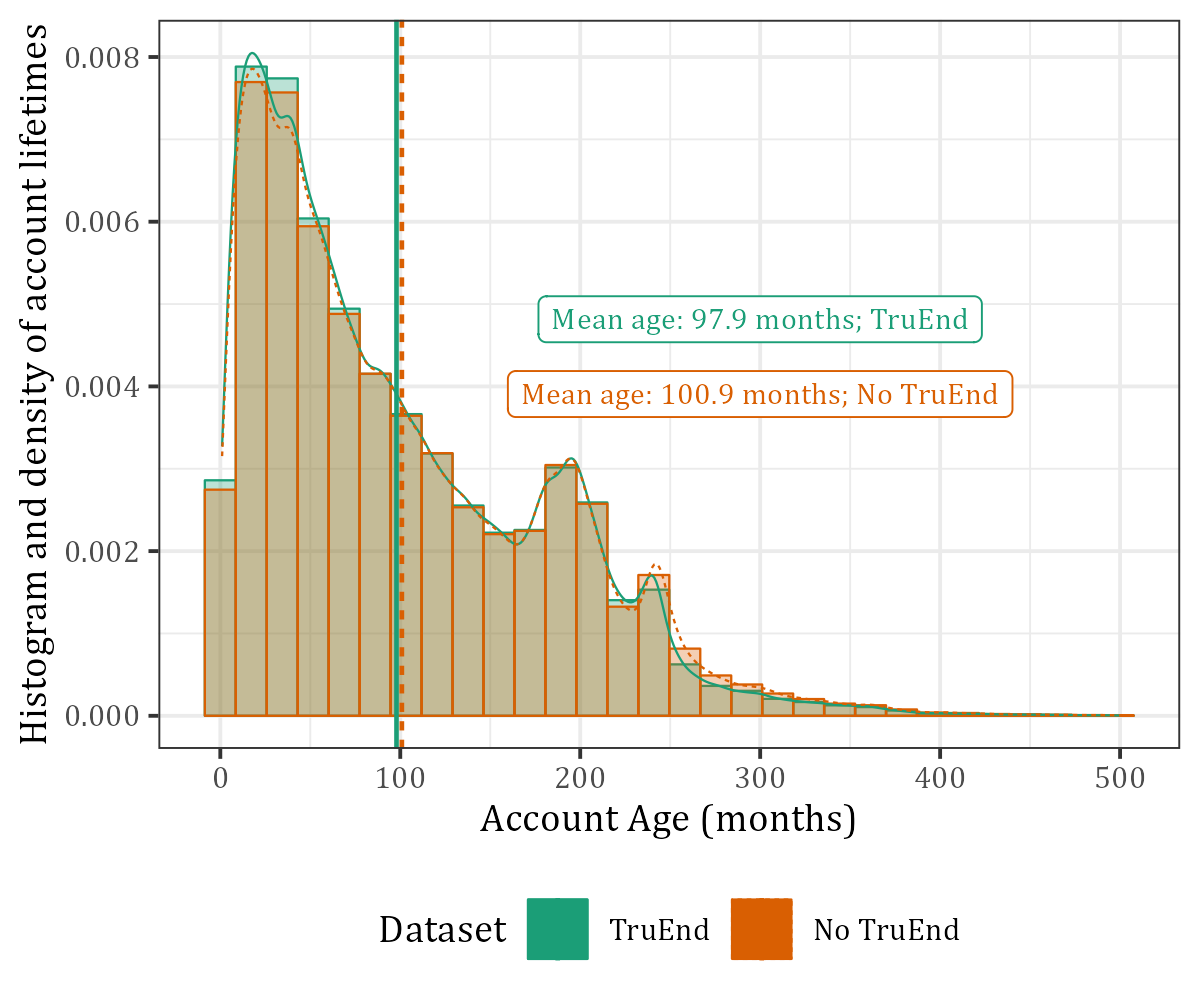}\label{fig:Impact_Densities_Age}
\end{subfigure} 
\begin{subfigure}[b]{0.49\textwidth}
    \caption{Realised loss rates (write-offs)}
    \centering\includegraphics[width=1\linewidth,height=0.3\textheight]{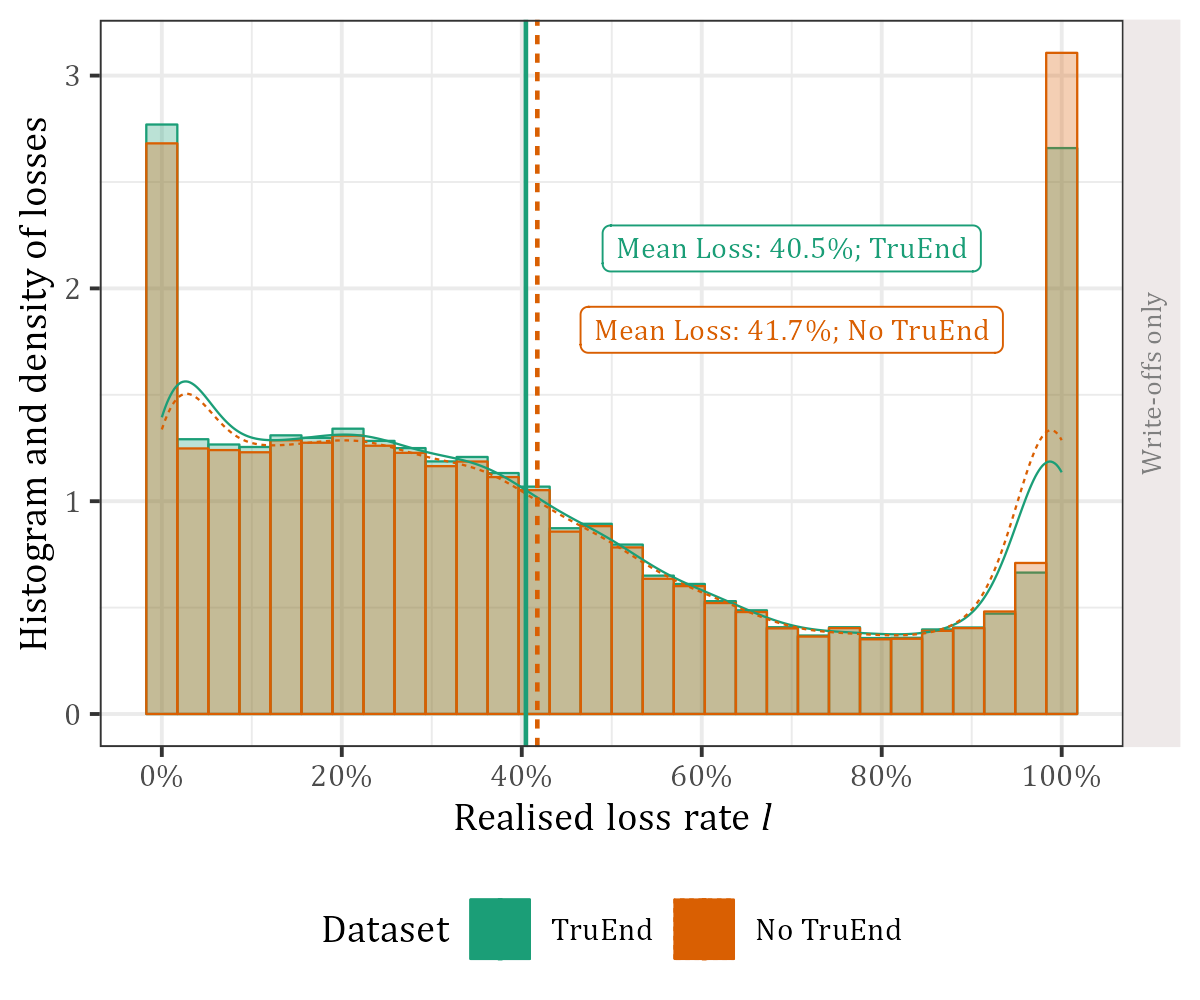}\label{fig:Impact_Densities_LGD}
\end{subfigure}%
\hfill%
\caption{A distributional analysis in comparing the histograms and densities of certain quantities in the mortgage dataset, before and after discarding the isolated TZB-cases using the optimal $b^*$-policy. Sample means are overlaid, showing a 3-month reduction in the average account age in \textbf{(a)}; and a 1.2\% point reduction in the average realised loss in \textbf{(b)}.}\label{fig:Impact_Densities}
\end{figure}

Having excluded the identified TZB-periods using this $b^*$-policy, we examine the impact thereof using a comparative distributional analysis. In particular, the empirical probability distribution of a certain quantity is analysed against that of its `predecessor'; i.e., the true distribution $f_\mathrm{T}$ that rightfully excludes TZB-periods vs the preceding false distribution $f_\mathrm{F}$ that still includes TZB-periods.
By virtue of shortening some account histories, we fully expect account ages $a$ to be affected, which should reflect in the distributional shapes of $f_\mathrm{F}$ and $f_\mathrm{T}$ across $a$. In fact, \autoref{fig:Impact_Densities_Age} shows that $f_\mathrm{F}(a)<f_\mathrm{T}(a)$ for smaller ages $a<80$ months, which reverses to $f_\mathrm{F}(a)>f_\mathrm{T}(a)$ for larger ages $a>240$. This change can also be described as a clockwise "distributional tilt" in the shape of $f_\mathrm{T}$, relative to that of $f_\mathrm{F}$. The overall mean age reduces by 3 months to 97.9 when applying this $b^*$-policy, even though its effect is much more pronounced for terminated accounts -- in which case, the mean age reduces by about 18 months.
Moreover, we obtain a similar result for the realised loss rates $l$, as calculated from written-off accounts using the workout-method with observed cash flows; see \citet{gurtler2013LGD} and \citet[\S 10]{baesens2016credit}. Specifically,  $f_\mathrm{T}$ also exhibits a clockwise distributional tilt relative to $f_\mathrm{F}$ across all $l$, as shown in \autoref{fig:Impact_Densities_LGD}. The mean loss rate therefore decreases by 1.2\% points to 40.5\% when applying this $b^*$-policy. While seemingly small, this reduction can amount to many millions saved in credit losses for typical mortgage portfolios, purely as an innocuous data fix before conducting any intensive modelling.

One can also gauge the impact of imposing this $b^*$-policy by using survival analysis to model another aspect of the LGD risk parameter: the time until write-off $T$. This non-negative random variable $T$ represents the age of a default spell that either ends with a resolved outcome (write-off) or is left unresolved as right-censored. 
Then, consider the cumulative lifetime probability distribution $F(t)=\mathbb{P}(T\leq t)$ across some time horizon $t$; see \citet[pp.~1-15]{kleinbaum2012survival}. Its complement is the write-off survivor function $S(t)=1-F(t)=\mathbb{P}(T>t)$, which may be estimated using the well-known non-parametric Kaplan-Meier (KM) estimator $\hat{S}(t)$ from \citet{kaplan1958credit}. 
Regarding competing risks that preclude write-off, the cured outcome is grouped with right-censored observations in the interest of expediency and simplicity. However, by following this "latent risks" approach, the quantity $\hat{F}(t)=1-\hat{S}(t)$ can be overestimated, as discussed in \citet{gooley1999estimation}. That said, the effects of such overestimation should at least be consistent within a broader two-component comparative study, especially when both components $A$ and $B$ are equally overestimated in assessing the difference $A-B$; itself the main objective at present. We therefore deem such overestimation to be negligible in its effect on the ensuing comparative study.

\begin{figure}[ht!]
\centering
\begin{subfigure}[b]{0.49\textwidth}
    \caption{Early time period $t\in[0,60]$}
    \centering\includegraphics[width=1\linewidth,height=0.3\textheight]{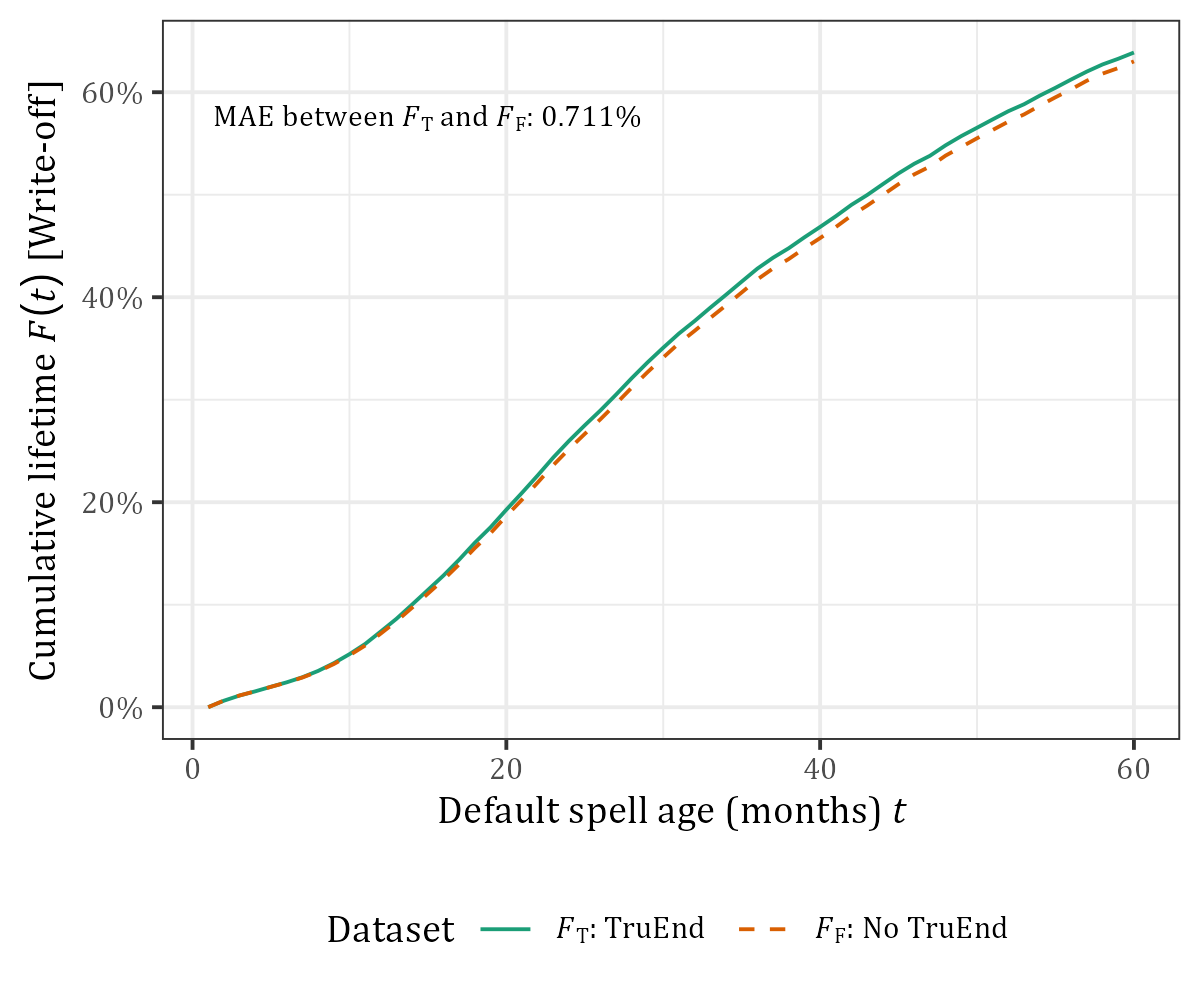}\label{fig:CumulLife_a}
\end{subfigure} 
\begin{subfigure}[b]{0.49\textwidth}
    \caption{Later time period $t\in[61,120]$}
    \centering\includegraphics[width=1\linewidth,height=0.3\textheight]{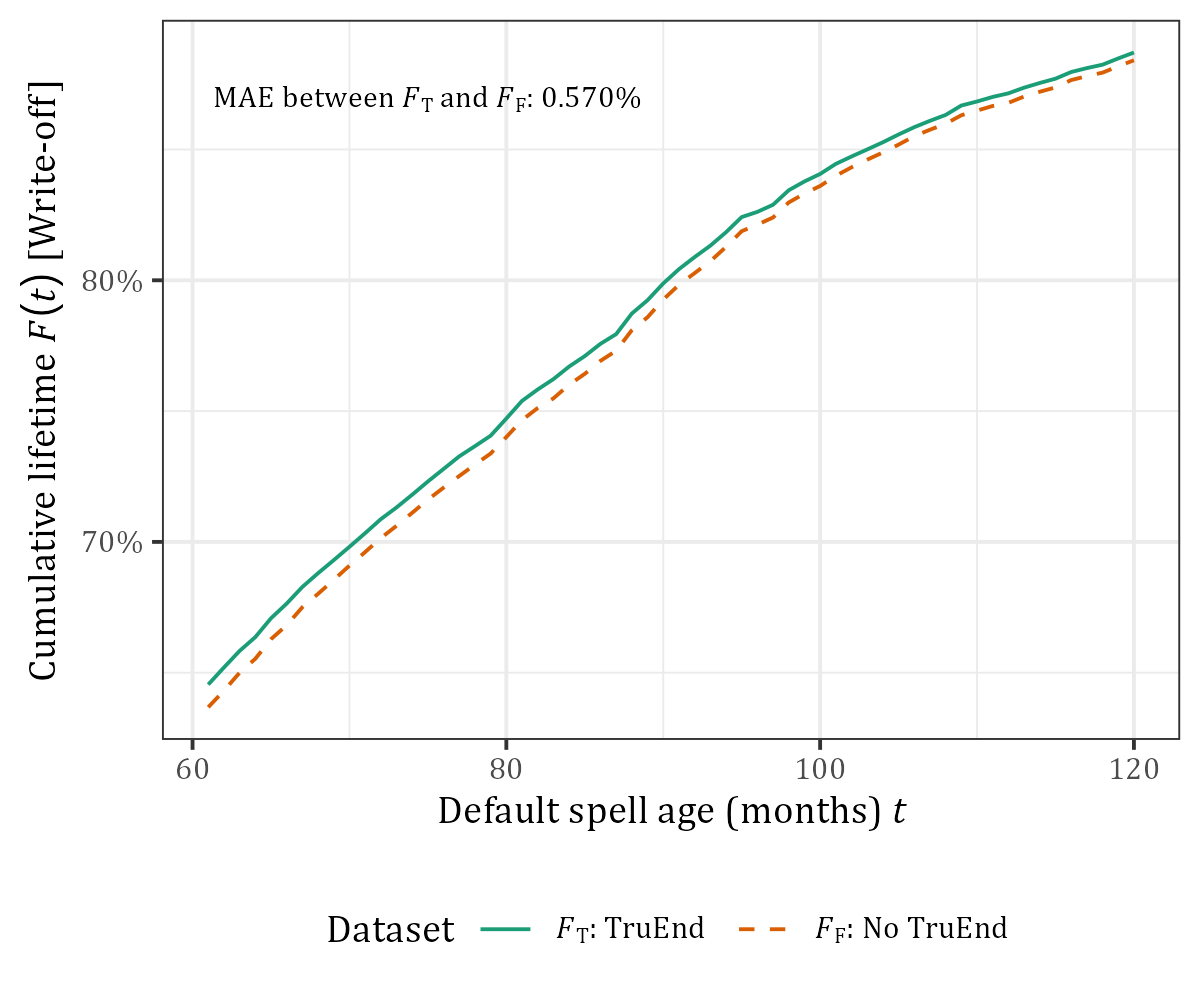}\label{fig:CumulLife_b}
\end{subfigure}%
\hfill%
\caption{Comparing the cumulative lifetime probability distribution $\hat{F}(t)$ over spell age $t$ until write-off, having applied the TruEnd-procedure ($F_\mathrm{T}$) vs its absence ($F_\mathrm{F}$). In \textbf{(a)}, $F(t)$ is graphed for the early period $t\in[0,60]$, whereas \textbf{(b)} shows the later period $\in[61,120]$. In summarising the discrepancy between the line graphs $A_t$ and $B_t$, the \textit{mean absolute error} (MAE) is calculated and overlaid.}\label{fig:CumulLife}
\end{figure}

\begin{figure}[ht!]
\centering
\begin{subfigure}[b]{0.49\textwidth}
    \caption{Early time period $t\in[0,60]$}
    \centering\includegraphics[width=1\linewidth,height=0.3\textheight]{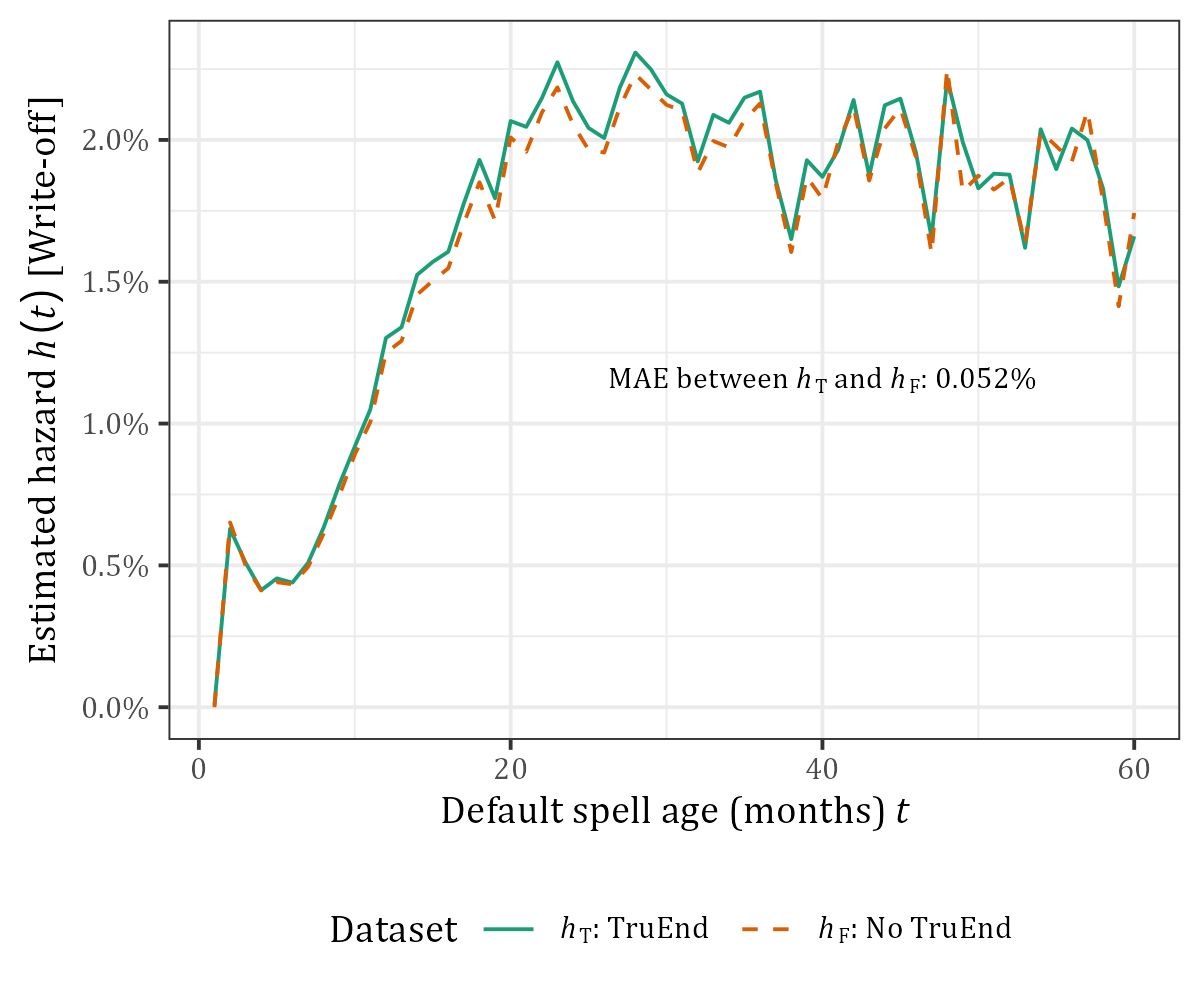}\label{fig:HazardFunc_a}
\end{subfigure} 
\begin{subfigure}[b]{0.49\textwidth}
    \caption{Later time period $t\in[61,120]$}
    \centering\includegraphics[width=1\linewidth,height=0.3\textheight]{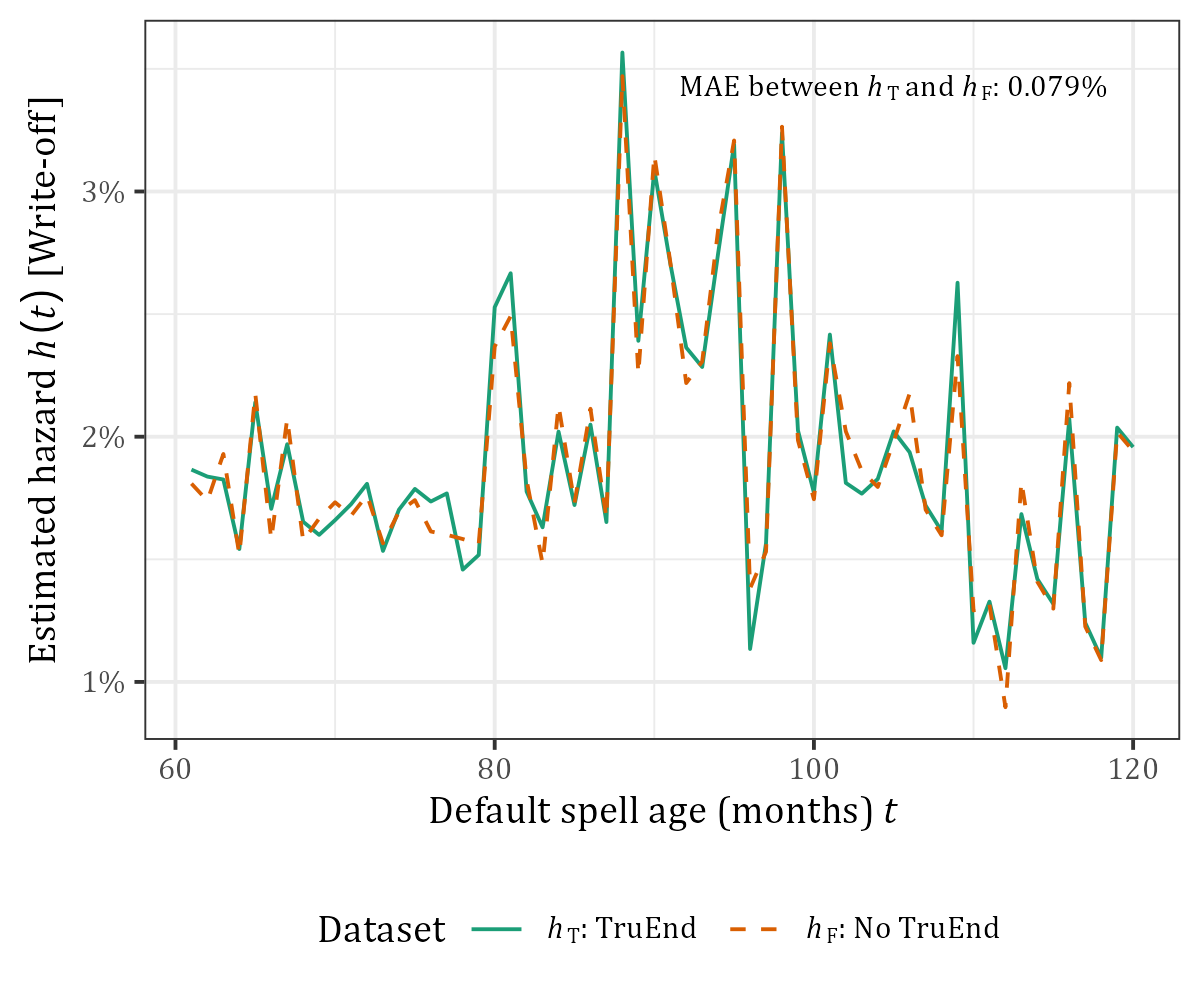}\label{fig:HazardFunc_b}
\end{subfigure}%
\hfill%
\caption{Comparing the hazard function $\hat{h}(t)$ over spell age $t$ until write-off, having applied the TruEnd-procedure ($h_\mathrm{T}$) vs its absence ($h_\mathrm{F}$); shown across two periods in (\textbf{a}) and (\textbf{b}). The graph design follows that of \autoref{fig:CumulLife}.}\label{fig:HazardFunc}
\end{figure}

We present the $\hat{F}(t)$-estimates in \autoref{fig:CumulLife} and compare the true distribution $F_\mathrm{T}(t)$ that rightfully excludes TZB-periods vs the false distribution $F_\mathrm{F}(t)$ that still includes TZB-periods; i.e., applying the TruEnd-procedure vs its absence. The difference between both cumulative distributions seems to grow up to a certain time point $t=84$, beyond which the difference gradually subsides again. 
The slightly larger lifetime write-off probabilities $F_\mathrm{T}(t)>F_\mathrm{F}(t)$ stem directly from the act of excluding TZB-periods, thereby moving risk events to earlier times during default spells. 
This result is corroborated in \autoref{fig:HazardFunc} wherein we provide estimates of the discrete-time hazard rate $\hat{h}(t)=\hat{f}(t) / \hat{S}(t-1)$, where $\hat{f}$ is the estimated density function of $\hat{F}$ that represents the frequency of failures per unit of time. The write-off hazard $\hat{h}(t)$ therefore denotes the conditional write-off probability of an account during $(t-1,t]$, having survived at least up to $t-1$. Accordingly, estimates of the true hazard distribution $h_\mathrm{T}(t)$ that rightfully exclude TZB-periods are compared to those of the false hazard distribution $h_\mathrm{F}(t)$ that still include TZB-periods. Again, we observe slightly larger write-off hazards $h_\mathrm{T}(t)\geq h_\mathrm{F}(t)$ at earlier times $t\leq 84$, though these differences grow smaller (and even reverse) at later times.
While write-off risk evidently increases when correcting the timing of the underlying risk events, the realised loss rates can decrease by virtue of shorter default spells, as previously shown in \autoref{fig:Impact_Densities}.
Lastly, default spells longer than 10 years are increasingly rare and of dubious practical value. We have therefore limited the (graphical) survival analyses of both $\hat{F}(t)$ and $\hat{h}(t)$ in \crefrange{fig:CumulLife}{fig:HazardFunc} to a maximum time in default of 10 years, without necessarily excluding those longer-aged default spells.

\section{Conclusion}
\label{sec:conclusion}

In credit data, an account's observed repayment history can incorrectly include a lengthy sequence of zero (or very small) month-end balances towards its end, often caused by operational or system failures that delay  the account's timely closure. This phenomenon, called \textit{trailing zero-valued balances} (TZB), poses a significant modelling challenge since it distorts the true but unobservable endpoints of affected accounts.
As a form of measurement error, these TZB-cases can ruin the timing of certain risk events such as write-off or early settlement. When training LGD-models from affected data, TZB-cases become particularly pernicious to the resulting loss predictions. The excess TZB-history can artificially prolong the period over which cash flows are discounted, thereby inflating the calculated loss percentage even before modelling such realised losses. Moreover, inaccurate timings of risk events in training data can compromise the subsequent predictions of any time-to-event model, including survival analysis. Therefore, the failure to identify and treat these TZB-cases will invite model risk, which can erode trust in all statistical models derived from such inaccurate data.

The remedy is intuitively simple: remove the `cancerous' TZB-affected part within the history of a loan, thereby moving its endpoint earlier. 
However, finding the true endpoint for each loan is difficult at scale, especially given the typically large sizes of credit datasets in retail lending. An automated procedure of sorts is clearly needed, though which would then need instruction on identifying these TZB-cases; itself a non-trivial task. Isolating such trails of diminutive balances will depend on the exact definition of a "small or immaterial" balance, which should itself be neither too small nor too large, lest we either retain false history or discard credible history.
Determining the `best' definition $b^*$ amidst many candidates $b$ would require an optimisation approach that embeds the underlying costs/benefits of each $b$-value. Such an approach would also need to contend with the uncertainty of account balances over time; i.e., the potential for settlement and bouts of loan delinquency, which may even end in write-off.
Moreover, the approach would need to find a single small-balance definition that can apply to the entire portfolio (or homogeneous segment therein), thereby exerting appropriate control while achieving consistency and simplicity.

We contribute such a data-driven optimisation approach that can find the best small-balance policy $b^*$ towards identifying and discarding latent TZB-periods within a given loan portfolio.
Our self-styled \textit{TruEnd-procedure} includes a bespoke objective function $f$ that is statistically evaluated across the histories of all constituent accounts. This $f$ relies on two opposing counter-measures, $M_2$ and $M_1$. Both measures give the mean balance of an account, albeit calculated over two different periods that respectively predate or succeed the supposed true endpoint. When optimising $f$ using a value iteration technique, the best $b^*$-policy effectively yields those individual endpoints within affected loan histories that simultaneously maximises $M_2$ and minimises $M_1$ on average.
Some of the novelty in these measures arises from their deliberate exploitation of the arithmetic mean's known weakness to outliers. Having identified the true endpoints using this $b^*$-policy, the excess histories thereafter constitute TZB-periods that can be discarded, thereby correcting the timing of account termination.

In demonstrating the TruEnd-procedure, we test a wide range of candidate policies $b$ in defining "small balances", having used two large datasets from the South African credit market: residential mortgages and credit card accounts.
Not only do we find the best $b^*$-policy within each dataset, but our results also reveal a small optimal region of $b$-values clustered around this $b^*$-value.
The optimisation results $f(b)$ are strikingly similar for those $b$-values within this region, which suggests that any one of them can practically serve as a policy definition for "small balances". 
However, the function $f$ reacts ever more drastically to larger values of $b$ outside of this region, whilst noting that such larger $b$-values would increasingly curtail the individual loan histories to their detriment. As $b$ grows larger beyond a certain point, the notion of cutting away more and more of the `healthy' parts of these histories should rightfully be penalised accordingly. Therefore, the associated and demonstrably severe reaction in $f$ serves as reassurance on the overall and intuitive soundness of the TruEnd-procedure in definitively disqualifying those larger $b$-values.
Lastly, we repeatedly applied the TruEnd-procedure within a broader Monte Carlo setup, which allowed us to inspect some of its statistical properties across many subsamples of mortgage data. Our results show that the optima from the procedure do not vary significantly, which attests to the stability and robustness of the TruEnd-procedure.

Aside from finding the optimal $b^*$-policy, TZB-affected accounts are remarkably prevalent within both datasets across all $b$-values. In particular, TZB-prevalence ranges from 15-28\% of mortgage accounts, and 17.5-40\% of credit card accounts. When applying any of these $b$-policies within the mortgage dataset, the isolated (and discarded) TZB-periods have significant durations in that loan histories are now shortened by about 14-19 months on average, depending on the exact $b$-value. Evidently, TZB-cases are both highly prevalent and material in this portfolio, which could not have been detected (or even quantified) if not for applying our TruEnd-procedure.
At the optimal $b^*$-policy of ZAR 300, we find that 23\% of accounts are TZB-cases with excess histories that span about 18 months, during which time the mean TZB-balance is practically zero-valued at ZAR 7.45. 
Having discarded the identified TZB-periods, the timing of account termination is therefore corrected, thereby reducing the mean loss rate of write-offs by 1.2\% points; a material result for any LGD-model.
Using survival analysis on defaulted loans, we further interrogate this result by modelling and comparing the now-corrected time to write-off versus its uncorrected counterpart. As expected, the write-off hazard generally increases at earlier durations within default spells, chiefly due to risk events now occurring slightly earlier than before.
Both the timing and severity of risk events are therefore made more accurate. By implication, the resulting LGD-distribution yields lower and less-biased empirical estimates of credit losses for defaulted loans.
Ultimately, we believe that this work can safely salvage a dataset that is otherwise corrupted by TZB-cases, which demonstrates the sound pursuit of innovation during data preparation.

Regarding limitations, the TruEnd-procedure currently uses monthly account balances $\boldsymbol{B}$ as the control variable in screening candidate thresholds $b$, respective to the domain of $\boldsymbol{B}$. However, these currency-denominated amounts are certainly susceptible to the erosive effects of inflation, especially over longer sampling periods. This susceptibility can detract from finding a single ideal threshold $b^*$ that should apply to all balances across time. Future research can therefore explore alternative controls such as the balance-to-principal ratio of an account over time, whose percentage-valued bounds can circumvent the aforementioned susceptibility to inflation.
In addition, future researchers can refactor the TruEnd-procedure towards using multiple control variables, e.g., both balance and balance-to-principal. Using multiple controls may afford even greater granularity in the search for optimal thresholds, though the added complexity thereof may become questionable.
Furthermore, our work can be extended by using a segmentation scheme in first partitioning the underlying dataset before running the optimisation step. Instead of yielding a single portfolio-wide threshold $b^*$, such future work can focus on finding an optimised threshold $b^*_s$ for each homogeneous segment $s$ of accounts.

It might also be worthwhile to examine more rigorously the effect on our results when varying the procedure's one parameter $(\tau)$, which was expediently fixed in this study to $\tau=\{6,12\}$ months. While our provisional results show little difference between either $\tau=6$ or $\tau=12$, future work may uncover greater sensitivity in $M_2$ for certain ranges of $\tau$.
Furthermore, the simplistic way in which our procedure is calibrated to a portfolio may warrant further research. Other ways of setting the weight $w$, which scales the effect of $M_2$ downward, may refine the optimisation results even further.
Lastly, other researchers could explore various techniques from the \textit{change point detection} (CPD) literature, as reviewed by \citet{aminikhanghahi2017survey}, in solving the same type of problem. Similar to the TruEnd-procedure, the common goal of these CPD-techniques is to detect a change point(s) in time series data beyond which the probability distribution has abruptly changed; i.e., TZB-periods. However, most CPD-techniques focus on screening a single time series for abrupt changes, whereas our problem context contains hundreds of thousands of very different time series (or loan histories). Future work could therefore focus on either scaling these CPD-techniques to our problem context, or explore our TruEnd-procedure as a new type of offline nonparametric CPD-technique within other domains.




\singlespacing
\printbibliography 

@phdthesis{botha2021Proc
,	author	= {Botha, Arno}
,	title	= {A procedure for loss-optimising the timing of loan recovery under uncertainty}
,	school	= {University of Pretoria}
,	year	= {2021}
,   doi     = {10.13140/RG.2.2.12015.30888/1}
}

@article{botha2020LROD
,	author	= {Botha, Arno and Beyers, Conrad and De Villiers, Pieter}
,	title	= {Simulation-based optimisation of the timing of loan recovery across different portfolios}
,	journal	= {Expert Systems with Applications}
,	year	= {2021}
,   volume  = {177}
,   doi     = {10.1016/j.eswa.2021.114878}
,   url     = {https://arxiv.org/abs/2009.11064}
}

@article{botha2020LROD_empirical
,	author	= {Botha, Arno and Beyers, Conrad and De Villiers, Pieter}
,	title	= {The loss optimisation of loan recovery decision times using forecast cash flows}
,	journal	= {Journal of Credit Risk}
,	year	= {2022}
,   doi     = {10.21314/JCR.2020.275}
,   url     = {https://arxiv.org/abs/2010.05601}
,   urldate = {2020-10-12}
}

@misc{botha2024sourcecode,
    title   = {{The TruEnd-procedure [source code] for treating trailing zero-valued balances in credit data}},
    author  = {Botha, Arno and Bester, Roelinde},
    year    = {2024},
    doi     = {10.5281/zenodo.10974650},
    url     = {https://doi.org/10.5281/zenodo.10974650},
    version = {1.1},
    publisher = {Zenodo}
}

@book{kleinbaum2012survival
,	author	= {Kleinbaum, David G and Klein, Mitchel}
,	title	= {Survival analysis: a self-learning text}
,	publisher	= {Springer}
,	year	= {2012}
,	edition	= {3}
,	doi	= {10.1007/978-1-4419-6646-9}
,	address	= {New York}
}

@book{james2013introduction
,	author	= {James, Gareth and Witten, Daniela and Hastie, Trevor and Tibshirani, Robert}
,	title	= {An introduction to statistical learning: with applications in {R}}
,	publisher	= {Springer}
,	year	= {2013}
,	address	= {New York}
,	doi	= {10.1007/978-1-4614-7138-7}
}

@book{baesens2016credit
,	author	= {Baesens, Bart and Rösch, Daniel and Scheule, Harald}
,	title	= {Credit Risk Analytics: Measurement Techniques, Applications, and Examples in {SAS}}
,	publisher	= {John Wiley \& Sons}
,	year	= {2016}
,	address	= {Hoboken, New Jersey}
}

@book{finlay2010book,
  title={The Management of Consumer Credit: Theory and Practice},
  edition={2},
  author={Finlay, Steven},
  year={2010},
  publisher={Palgrave Macmillan},
  address={Hampshire, UK}
}

@book{thomas2009consumer
,	author	= {Thomas, Lyn C}
,	title	= {Consumer Credit Models: Pricing, Profit and Portfolios}
,	publisher	= {Oxford University Press}
,	year	= {2009}
,	address	= {New York}
,	doi	= {10.1093/acprof:oso/9780199232130.001.1}
}

@book{boyd2004convex,
  title={Convex optimization},
  author={Boyd, Stephen and Vandenberghe, Lieven},
  year={2004},
  publisher={Cambridge University Press}
}

@manual{ifrs9_2014
,	author	= {IASB}
,	title	= {International Financial Reporting Standard ({IFRS}) 9: Financial Instruments}
,	organization = {IFRS Foundation: International Accounting Standards Board (IASB)}
,	year	= {2014}
,	address	= {London}
,	url	= {https://www.ifrs.org/issued-standards/list-of-standards/ifrs-9-financial-instruments/}
}

@manual{pra2023PSmodelrisk
, title = {{PS6/23: Model risk management principles for banks}}
, author = {PRA}
, organization = {Bank of England, Prudential Regulation Authority (PRA)}
, year = {2023}
, url = {https://www.bankofengland.co.uk/prudential-regulation/publication/2023/may/model-risk-management-principles-for-banks}
, address = {United Kingdom}
}

@manual{pra2023SSmodelriskprinciples
, title = {{SS1/23: Model risk management principles for banks}}
, author = {PRA}
, organization = {Bank of England, Prudential Regulation Authority (PRA)}
, year = {2023}
, url = {https://www.bankofengland.co.uk/prudential-regulation/publication/2023/may/model-risk-management-principles-for-banks-ss}
, address = {United Kingdom}
}

@manual{sarb2022g9
, title = {G9/2022: Matters related to the credit risk models of banks}
, author = {SARB}
, year = {2022}
, url = {https://www.resbank.co.za/en/home/publications/publication-detail-pages/prudential-authority/pa-deposit-takers/banks-guidance-notes/2022/G9-2022-Matters-related-to-the-credit-risk-models-of-banks}
, organization = {South African Reserve Bank (SARB):}
}

@article{singer1993time,
  title={It’s about time: Using discrete-time survival analysis to study duration and the timing of events},
  author={Singer, Judith D and Willett, John B},
  journal={Journal of Educational Statistics},
  volume={18},
  number={2},
  pages={155--195},
  year={1993},
  DOI = {10.2307/1165085}
}

@article{schober2018survival,
  title={Survival analysis and interpretation of time-to-event data: the tortoise and the hare},
  author={Schober, Patrick and Vetter, Thomas R},
  journal={Anesthesia and Analgesia},
  volume={127},
  number={3},
  pages={792-798},
  year={2018},
  publisher={Wolters Kluwer Health},
  doi={10.1213/ANE.0000000000003653}
}

@article{kaplan1958credit,
  URL = {http://www.jstor.org/stable/2281868},
 author = {E. L. Kaplan and Paul Meier},
 journal = {Journal of the American Statistical Association},
 number = {282},
 pages = {457--481},
 publisher = {[American Statistical Association, Taylor & Francis, Ltd.]},
 title = {Nonparametric Estimation from Incomplete Observations},
 urldate = {2023-03-02},
 volume = {53},
 year = {1958}
}

@article{kartsonaki2016survival,
  title={Survival analysis},
  author={Kartsonaki, Christiana},
  journal={Diagnostic Histopathology},
  volume={22},
  number={7},
  pages={263--270},
  year={2016},
  publisher={Elsevier},
  doi={10.1016/j.mpdhp.2016.06.005}
}

@article{dirick2017time
,	author	= {Dirick, Lore and Claeskens, Gerda and Baesens, Bart}
,	title	= {Time to default in credit scoring using survival analysis: a benchmark study}
,	journal	= {Journal of the Operational Research Society}
,	year	= {2017}
,	volume	= {68}
,	number	= {6}
,	pages	= {652--665}
,	publisher	= {Taylor \\& Francis}
,	doi	= {10.1057/s41274-016-0128-9;}
}

@inproceedings{narain1992credit,
  author={Narain, B},
  title={Survival analysis and the credit granting decision},
  booktitle={Credit Scoring and Credit Control},
  editor={Thomas, Lyn C and Crook, Jonathan N and Edelman, David B},
  pages = {109-121},
  year={1992},
  publisher={OUP, Oxford, UK}
}

@article{banasik1999not
,	author	= {Banasik, John and Crook, Jonathan N and Thomas, Lyn C}
,	title	= {Not if but when will borrowers default}
,	journal	= {Journal of the Operational Research Society}
,	year	= {1999}
,	pages	= {1185--1190}
,	publisher	= {JSTOR}
,	doi	= {10.1057/palgrave.jors.2600851}
,	volume	= {50}
,	number	= {12}
}

@article{skoglund2017, 
    title={Credit risk term-structures for Lifetime Impairment Forecasting: A practical guide}, 
    journal={Journal of Risk Management in Financial Institutions},
    volume	= {10},
    number	= {2},
    pages	= {177--195},
    url = {https://www.econbiz.de/Record/credit-risk-term-structures-for-lifetime-impairment-forecasting-a-practical-guide-skoglund-jimmy/10011670671},
    author={Skoglund, Jimmy}, year={2017}
}

@article{novotny2016
,	author	= {Novotny-Farkas, Zolt\'an}
,	title	= {The interaction of the {IFRS} 9 expected loss approach with supervisory rules and implications for financial stability}
,	journal	= {Accounting in Europe}
,	year	= {2016}
,	volume	= {13}
,	number	= {2}
,	pages	= {197--227}
,	doi	= {10.1080/17449480.2016.1210180}
}

@article{schuermann2004we
, author = {Schuermann, Til}
, title = {What do we know about Loss Given Default?}
, year = {2004}
, journal = {Wharton Financial Institutions Centre Working Paper}
,   howpublished = {Availaible at SSRN:}
, address = {London, UK}
,   doi     = {10.2139/ssrn.525702}
}

@article{calabrese2010bank
,	author	= {Calabrese, Raffaella and Zenga, Michele}
,	title	= {Bank loan recovery rates: Measuring and nonparametric density estimation}
,	journal	= {Journal of Banking \& Finance}
,	year	= {2010}
,	volume	= {34}
,	number	= {5}
,	pages	= {903--911}
,	publisher	= {Elsevier}
,	doi	= {10.1016/j.jbankfin.2009.10.001}
}

@article{gurtler2013LGD
,	author	= {G\"urtler, Marc and Hibbeln, Martin}
,	title	= {Improvements in loss given default forecasts for bank loans}
,	journal	= {Journal of Banking \& Finance}
,	year	= {2013}
,	volume	= {37}
,	number	= {7}
,	pages	= {2354--2366}
,	publisher	= {Elsevier}
,	doi	= {10.1016/j.jbankfin.2013.01.031}
}

@article{stepanova2002survival
,	author	= {Stepanova, Maria and Thomas, Lyn}
,	title	= {Survival analysis methods for personal loan data}
,	journal	= {Operations Research}
,	year	= {2002}
,	volume	= {50}
,	number	= {2}
,	pages	= {277--289}
,	publisher	= {INFORMS}
,	doi	= {10.1287/opre.50.2.277.426}
}

@article{dejongh_2017, 
    author={De Jongh, Pieter J. and Larney, Janette and Mare, Eben and Van Vuuren, Gary W. and Verster, Tanja}, 
    year={2017},
    title={A proposed best practice model validation framework for Banks}, 
    volume={20},
    number={1}, 
    journal={{South African Journal of Economic and Management Sciences}},
    DOI={10.4102/sajems.v20i1.1490}
}

@article{bellotti2013, 
    title={Forecasting and stress testing credit card default using dynamic models}, 
    volume={29}, 
    DOI={10.1016/j.ijforecast.2013.04.003}, 
    number={4}, 
    journal={International Journal of Forecasting}, 
    author={Bellotti, Tony and Crook, Jonathan}, 
    year={2013}, 
    pages={563–574}
}

@article{crook2010dynamic, 
  title={Time Varying and dynamic models for default risk in consumer loans}, 
  volume={173}, 
  number={2}, 
  journal={Journal of the Royal Statistical Society: Series A (Statistics in Society)}, 
  author={Crook, Jonathan and Bellotti, Tony}, 
  year={2010}, 
  pages={283–305},
  DOI={10.1111/j.1467-985x.2009.00617.x}
}

@article{bellotti2009macro
,	author	= {Bellotti, Tony and Crook, Jonathan}
,	title	= {Credit scoring with macroeconomic variables using survival analysis}
,	journal	= {Journal of the Operational Research Society}
,	year	= {2009}
,	volume	= {60}
,	number	= {12}
,	pages	= {1699--1707}
,	publisher	= {Nature Publishing Group}
,	doi	= {10.1057/jors.2008.130}
}

@article{bellotti2014stresstesting
,	author	= {Bellotti, Tony and Crook, Jonathan}
,	title	= {Retail credit stress testing using a discrete hazard model with macroeconomic factors}
,	journal	= {Journal of the Operational Research Society}
,	year	= {2014}
,	volume	= {65}
,	number	= {3}
,	pages	= {340--350}
,	publisher	= {Taylor \\& Francis}
,	doi	= {10.1057/jors.2013.91}
}

@article{wood2017addressing
,	author	= {Wood, Richard and Powell, David}
,	title	= {Addressing probationary period within a competing risks survival model for retail mortgage loss given default}
,	journal	= {Journal of Credit Risk}
,	year	= {2017}
,	volume	= {13}
,	number	= {3}
,	doi	= {10.21314/JCR.2017.228}
}

@article{zhang2012comparisons,
  title={Comparisons of linear regression and survival analysis using single and mixture distributions approaches in modelling {LGD}},
  author={Zhang, Jie and Thomas, Lyn C},
  journal={International Journal of Forecasting},
  volume={28},
  number={1},
  pages={204--215},
  year={2012},
  publisher={Elsevier},
  doi={10.1016/j.ijforecast.2010.06.002}
}

@article{witzany2012survival
,	author	= {Witzany, Jiri and Rychnovsky, Michal and Charamza, Pavel}
,	title	= {Survival analysis in {LGD} modeling}
,	journal	= {European Financial and Accounting Journal}
,	year	= {2012}
,	volume	= {7}
,	doi	= {10.18267/j.efaj.12}
,	number	= {1}
,	pages	= {6-27}
,  doi={10.18267/j.efaj.12}
}

@article{larney2023modelling,
  title={Modelling the Time to Write-Off of Non-Performing Loans Using a Promotion Time Cure Model with Parametric Frailty},
  author={Larney, Janette and Allison, James Samuel and Grobler, Gerrit Lodewicus and Smuts, Marius},
  journal={Mathematics},
  volume={11},
  number={10},
  pages={2228},
  year={2023},
  publisher={MDPI},
  doi={10.3390/math11102228}
}

@article{joubert2018making,
  title={Making use of survival analysis to indirectly model loss given default},
  author={Joubert, Morne and Verster, Tanja and Raubenheimer, Helgard},
  journal={ORiON},
  volume={34},
  number={2},
  pages={107--132},
  year={2018},
  publisher={ORSSA},
  doi={10.5784/34-2-588}
}

@article{gooley1999estimation,
  title={Estimation of failure probabilities in the presence of competing risks: new representations of old estimators},
  author={Gooley, Ted A and Leisenring, Wendy and Crowley, John and Storer, Barry E},
  journal={Statistics in medicine},
  volume={18},
  number={6},
  pages={695--706},
  year={1999},
  publisher={Wiley Online Library},
doi={10.1002/(sici)1097-0258(19990330)18:6<695::aid-sim60>3.0.co;2-o}
}

@article{aminikhanghahi2017survey,
  title={A survey of methods for time series change point detection},
  author={Aminikhanghahi, Samaneh and Cook, Diane J},
  journal={Knowledge and information systems},
  volume={51},
  number={2},
  pages={339--367},
  year={2017},
  publisher={Springer},
  doi={10.1007/s10115-016-0987-z}
}
\onehalfspacing



\end{document}